\documentclass[11pt,a4paper]{article}
\usepackage{jheppub}
    \makeatletter
\def\@fpheader{\relax}
\makeatother
\usepackage{mathtools}
\usepackage{graphicx}
	\graphicspath{ {images/} }
\usepackage{blindtext}	
\usepackage{amsmath,amssymb,amsthm}
\usepackage{latexsym,graphicx,color}%,subfigure}
\usepackage{enumerate}
\usepackage{color}

\def\half{\frac{1}{2}}
\def\a{\alpha}

\def\half{\frac{1}{2}}
\usepackage{hyperref}

\newcommand{\D}{\mathcal{D}}
\newcommand{\p}{\partial}
\newcommand{\Tr}{{\rm Tr}}

\newcommand{\bea}{\begin{eqnarray}}
\newcommand{\eea}{\end{eqnarray}}
\def\Tr{ \hbox{\rm Tr}}

\def\l{\left}
\def\r{\right}

\def\c{\mathcal{C}}
\title{{\bf Aharonov-Bohm defects
}}
\author{Chandrasekhar Chatterjee$^{a}$} 
\author{Muneto Nitta$^{b}$}
 \affiliation{Department of Physics $\&$ Research and Education Center for Natural Sciences,\\ Keio University, Hiyoshi 4-1-1, Yokohama, Kanagawa 223-8521, Japan}
 \emailAdd{chandra.chttrj@gmail.com$^{a}$, chandra@phys-h.keio.ac.jp$^{a}$}
 \emailAdd{nitta(at)phys-h.keio.ac.jp$^{b}$}

\date{\today}

\abstract{
We discuss what happens when a field receiving an 
Aharonov-Bohm (AB) 
phase develops a vacuum expectation value (VEV), 
with an example of an Alice string
in a $U(1) \times SU(2)$ gauge theory coupled with 
complex triplet scalar fields. 
We introduce scalar fields belonging to 
the doublet representation of $SU(2)$,
charged or chargeless under the $U(1)$ gauge symmetry, 
that receives an AB phase around the Alice string. 
When the doublet develops a VEV, the Alice string 
turns to a global string in the absence of 
the interaction depending on the relative phase between 
the doublet and triplet, 
while, in the presence of such an interaction, the Alice string 
is confined by a soliton or domain wall 
and therefore the spontaneous breaking of a spatial rotation around the string is accompanied.
We call such an object induced by an AB phase as an ``AB defect'', 
and argue that such a phenomenon is ubiquitously appearing in various systems.
}
\begin{document}

\maketitle

\section{Introduction}

A gauge potential rather than a field strength (a magnetic or electric field) 
is not merely a mathematical object but a physical quantity, 
as manifested by the Aharonov-Bohm (AB) effect  \cite{Aharonov:1959fk}, 
which is a quantum mechanical effect occurring when a charged particle scatters from 
a solenoid with non-zero magnetic flux inside;  
Although both the magnetic and electric fields are zero everywhere outside the solenoid, 
a particle going around the solenoid is affected by the gauge potential 
and picks up a phase, 
resulting in a non-trivial differential scattering cross section. 
The AB effect was experimentally observed in seminar papers 
\cite{Tonomura:1982,Tonomura:1986} 
and then has been examined in various nano materials such as quantum dots.
Nowadays, studies of the AB effects are not only limited to materials 
but are also explored into various areas of physics, 
from particle physics, quantum field theory and string theory to cosmology.   
In cosmology, AB cosmic strings, {\it i.~e.}, 
cosmic strings (vortices) exhibiting AB effects were proposed \cite{Alford:1988sj}, 
and it was discussed that they experience frictions due to the AB effects  
 \cite{Vilenkin:1991zk,MarchRussell:1991az}.  
It was also suggested that AB cosmic strings may give a possible observational evidence of string theory 
\cite{Polchinski:2005bg,Ookouchi:2013gwa}.  
Non-Abelian vortices in supersymmetric gauge theory 
\cite{Hanany:2003hp,Auzzi:2003fs,Hanany:2004ea,Shifman:2004dr,Eto:2004rz,Gorsky:2004ad,Eto:2005yh,Eto:2006cx,Eto:2006db,Tong:2005un,Eto:2006pg,Shifman:2007ce,Shifman:2009zz}
exhibit AB effects \cite{Evslin:2013wka,Bolognesi:2015mpa,Bolognesi:2015ida} 
including non-Abelian generalization of AB effects \cite{Horvathy:1985jr},
once a part of flavor symmetry is gauged \cite{Konishi:2012eq}.
In dense QCD which may be relevant for cores of neutron stars,  
a color magnetic flux tube in the 2SC phase exhibits AB effects for quarks 
\cite{Alford:2010qf}, while  
 a non-Abelian vortex (color magnetic flux tube) in the color-flavor locked phase 
 \cite{Balachandran:2005ev,Nakano:2007dr,Eto:2009kg,Eto:2009bh,Eto:2009tr,Eto:2013hoa} 
also exhibits (electromagnetic) AB effects for charged particles
\cite{Chatterjee:2015lbf} 
as well as ${\mathbb Z}_3$ (color) AB effects for quarks  
 \cite{Cherman:2018jir,Chatterjee:2018nxe,Chatterjee:2019tbz,Hirono:2018fjr}.

An Alice string is one of strings exhibiting non-trivial AB phases. 
It is a kind of topological vortex which changes the sign of the charge of particles encircling around it \cite{Schwarz:1982ec, Kiskis:1978ed}. 
The conventional model admitting an Alice string is given by an $SO(3)$ gauge theory with 
scalar fields belonging to the fiveplet representation (traceless symmetric tensor),
in which the $SO(3)$ gauge symmetry is spontaneously broken down to $O(2)$.
The unbroken gauge group $O(2)$ is a subgroup inside the full symmetry group and becomes space dependent around the Alice string.
The $U(1)$ generator in $O(2)$ which we identify as the electromagnetism flips the sign after encircling once around the Alice string. This property generates an AB phase of charged particles. 
In the literature, these strings were discussed from purely field theoretical interests such 
 as non-local charge called a Cheshire charge, non-Abelian statistics and so on
 \cite{Alford:1990mk, Alford:1990ur, Preskill:1990bm,  Alford:1992yx, Bucher:1992bd, Bucher:1993jj, Lo:1993hp,Striet:2000bf,Benson:2004ue}. 
 A global analogue of Alice strings was discussed in the context of spinor Bose-Einstein condensates 
 \cite{Leonhardt:2000km,Ruostekoski:2003qx,Kobayashi:2011xb,Kawaguchi:2012ii}.
Analogues of Alice strings in string theory were also discussed 
\cite{Harvey:2007ab,Harvey:2008zz,Okada:2014wma}.
Recently, a Bogomol'nyi completion 
 \cite{Bogomolny:1975de,Prasad:1975kr} 
 for an Alice string was achieved in 
  a $U(1) \times SU(2)$ gauge theory coupled with complex triplet scalar fields,    
  and it was shown to be a half Bogomol'nyi-Prasad-Sommerfield (BPS) state 
  in supersymmetric theories \cite{Chatterjee:2017jsi,Chatterjee:2017hya}. 
  The point was that the fundamental group is $\pi_1(M) \simeq {\mathbb Z}$ unlike 
  the conventional case of $\pi_1(M) \simeq {\mathbb Z}_2$ where $M$ is a vacuum manifold. 
Since the energy is saturated by topological charges for BPS states, 
one can construct in principle multiple string configurations placed at arbitrary positions.
 This model was in fact a local version of global Alice strings in Refs.~\cite{Leonhardt:2000km,Ruostekoski:2003qx,Kobayashi:2011xb,Kawaguchi:2012ii}.
Recently, an Alice string was also studied in the context of axion cosmology 
\cite{Sato:2018nqy,Chatterjee:2019rch}, in which the $U(1)$ part 
in the full symmetry group $U(1) \times SU(2)$ 
is global making it to be a global (axion) string while the $SU(2)$ part is local.

In this paper, 
we discuss what happens when a field receiving a non-trivial AB 
phase develops a vacuum expectation value (VEV),  
with an example of an Alice string
in a $U(1) \times SU(2)$ gauge theory coupled with 
complex triplet scalar fields. 
We then introduce doublet scalar fields 
charged or chargeless under the $U(1)$ group, 
which receive non-trivial AB phases in the presence of an Alice string. 
We study the behavior of the Alice string when the doublet fields develop a VEV. 
We show that a soliton or domain wall attached to the Alice string 
is inevitably created  
in the presence of an interaction depending on the relative phase between the doublet and triplet,
and therefore the spontaneous breaking of a spatial rotation around the string is accompanied. 
We also show that in the absence of such an interaction, no soliton appears and 
the Alice string turns to a global string. 
We also find that the backreaction due to the existence of two condensates living together 
makes the magnetic flux inside the Alice string fractional. 

We should note that axion domain walls attached to an axion string 
(see Ref.~\cite{Kawasaki:2013ae} as a review) 
are not AB defects although configuration of string-wall composites 
\cite{Kibble:1982dd,Vilenkin:1982ks} 
themselves look similar to AB defects. 
Another example of an AB defect can be found in 
the Georgi-Machacek model \cite{Georgi:1985nv} 
proposed as a model beyond the standard model (SM), 
having three real triplet scalar fields and one doublet scalar field.
If the triplet VEVs are larger than the doublet VEV,
then a ${\mathbb Z}_2$ string is attached by a domain wall \cite{Chatterjee:2018znk}, 
as we will discuss in discussion.

This paper is organized as follows.
 In Sec.~\ref{sec:model}, we describe our model of an $SU(2)\times U(1)$ gauge theory with 
 one complex triplet and one doublet scalar fields. We consider two different interaction potentials corresponding to the two models for which the doublet field is charged or chargeless under the $U(1)$ gauge symmetry. In Sec.~\ref{sec:Alice-AB}, we briefly review Alice string in the $SU(2)\times U(1)$ gauge theory coupled with only complex triplet scalar fields. 
 Although BPS-ness is not necessary, we consider a BPS Alice string just for simplicity.
 We also present AB phases of the charged or chargeless doublet fields.  
 In Sec.~\ref{sec:model1}, we discuss the vacuum structure, a global Alice string with a fractional flux 
 and an Alice string attached by a soliton in  
 the first model for which the doublet field is charged. 
 In Sec.~\ref{sec:model2},
 we discuss the same in
 the second model for which the doublet field is chargeless.
 Sec.~\ref{sec:summary} is devoted to a summary and discussion. 
 We argue that the appearance of AB defects is ubiquitous in various field theoretical and condensed matter models.
In Appendix \ref{sec:appendix} we give a detailed explanation of our numerical method.

%%%%%%%%%%%%%%%%%%%%%%%%%%%%%%%%%%%%%%%%%%%%%
\section{The Models}\label{sec:model}
We start with an $SU(2)\times U(1)$ gauge theory coupled with 
one complex triplet scalar field $\Phi$ and one doublet scalar field $\Psi$. 
Matter contents are the same with the triplet Higgs model for beyond the SM, 
but we consider a different phase and different parameter region. 
However, we use terminology ``hypercharge'' for 
the $U(1)$ part and label it as $U(1)_Y$.
The hypercharge of  the triplet is fixed as $Y = 1$, 
and we discuss the two cases of the doublet fields carrying 
the hypercharge $Y = \half$ (as the same as the triplet Higgs model) and 
$Y=0$.

The Lagrangian is given by
\begin{eqnarray}
\mathcal{L}_{\rm kin} &=& -\frac{1}{4} \Tr F_{\mu\nu}F^{\mu\nu} - \frac{1}{4}f_{\mu\nu}f^{\mu\nu} + \Tr | D_\mu\Phi|^2 +  \left|\D_\mu\Psi\right|^2 - V(\Phi,\Psi)
\end{eqnarray}
where 
$
\D_\mu\Phi^\a  = \p_\mu\Phi^\a - ie_{} a_\mu \Phi + g \epsilon^{\alpha\beta\gamma}A_\mu^\beta\Phi^\gamma,\,
 \D_\mu\Psi = \p_\mu \Psi  - i\frac{e_\psi}{2} a_\mu \Psi - i A_\mu \Psi,
 F_{\mu\nu } = \p_\mu A_\nu  - \p_\nu A_\mu- i g [A_\mu, A_\nu],\, f_{\mu\nu } = \p_\mu a_\nu  - \p_\nu a_\mu
$.
Here $e_{}$ and $g$ are the coupling constants of $U(1)_Y$ and $SU(2)$ gauge fields, respectively, where $e_\psi$ takes values 
$\{e_{}, 0\}$ for the two different cases of the doublet field.
 The potential term is given by
\begin{eqnarray}
\label{potential}
 V(\Phi, \Psi) &=& V_\Phi (\Phi) + V_\Psi(\Psi) + V_{\rm int}(\Phi, \Psi).
\end{eqnarray} 
The potentials for the triplet and doublet fields, given by 
\begin{eqnarray}
 V_\Phi(\Phi) &=& \frac{\lambda_g}{4} \Tr[\Phi,\Phi^\dagger]^2 +  \frac{\lambda_e}{2}\left(\Tr \Phi\Phi^\dagger - 2 \xi^2\right)^2 , \label{eq:VPhi} \\
 V_\Psi(\Psi) &=&  M^2 \Psi^\dagger\Psi + \lambda_\psi \l(\Psi^\dagger\Psi\r)^2 ,
 \label{eq:VPsi}
\end{eqnarray}
respectively, are common for the two models, 
where
 $\lambda_g$ and $\lambda_e$ are two couplings of the triplet, 
  $\xi$ is a parameter giving a VEV of the triplet,  $M$ is the bare mass of the doublet,  $\lambda_\psi$ is the quartic coupling of the doublet field.
As for the interaction term
 $V_{\rm int}$
between the doublet and triplet fields, 
we consider 
\begin{eqnarray} 
 V_{\rm int}^{(1)}(\Phi, \Psi) &=& \mu^{(1)} \l( \Psi_c^\dagger \Phi^* \Psi +   \Psi^\dagger \Phi \Psi_c\r)  + 
 \lambda_1 \Psi^\dagger\Psi \Tr \l(\Phi^\dagger\Phi\r)+ \lambda_2 \Psi^\dagger\Phi^\dagger\Phi\Psi, \label{eq:int1} \\
 V_{\rm int}^{(2)}(\Phi, \Psi) &=& \mu^{(2)} \l| \Psi_c^\dagger \Phi^* \Psi\r|^2  + 
 \lambda_1 \Psi^\dagger\Psi \Tr \l(\Phi^\dagger\Phi\r) + \lambda_2 \Psi^\dagger\Phi^\dagger\Phi\Psi  \label{eq:int2}
\end{eqnarray}
for two models, 
where $\mu^{(1)}$ and $\mu^{(2)}$ are parameters of terms 
depending on the relative phase between the doublet and triplet fields, 
different for the two models, and 
$\lambda_1, \lambda_2$ are other couplings between these fields, common for the two models.
The charge conjugation of the doublet field
is defined as $\Psi_c = i\sigma^2 \Psi^*$. 

%%%%%%%%%%%%%%%%%%%%%%%%
\section{Alice string and Aharonov-Bohm phases around it}\label{sec:Alice-AB}

\subsection{Alice string solution: a review}
In this section, we  give a brief review  of BPS Alice strings 
  \cite{Chatterjee:2017jsi,Chatterjee:2017hya}
 in the absence of the doublet field.  So in this case we set all the scalar couplings are zero except $\lambda_g$ and $\lambda_e$. To construct a vortex solution we  write the static Hamiltonian as
\begin{eqnarray}
 \label{alice_Hamiltonian}
&&{H} = \int d^3x \left[\frac{1}{2} \Tr F_{ij} F^{ij} + \frac{1}{4} f_{ij}f^{ij} + \Tr | D_i\Phi|^2  
+  \frac{\lambda_g}{4} \Tr[\Phi,\Phi^\dagger]^2
+  \frac{\lambda_e}{2}\left(\Tr \Phi\Phi^\dagger  - 2 \xi^2\right)^2\right]. \nonumber\\
\end{eqnarray}
We  choose the vacuum expectation value of the field $\Phi$ as 
\begin{eqnarray}\label{phivac}
\langle\Phi\rangle_v = 2 \xi \tau^1,\;
\end{eqnarray}
with $\tau^1 = \half \sigma^1$. 
This vacuum is different from what in general used in the triplet Higgs model beyond the SM. 
This triplet vacuum breaks the gauge symmetry group spontaneously  as 
\begin{eqnarray}
\label{SSB1}
 G =  U(1)_Y \times SU(2)  \longrightarrow  
 H = \mathbb{Z}_2 \ltimes  U(1)_1 \simeq O(2) ,
\end{eqnarray}
where $\ltimes$ stands for a semi-direct product. Here in $U(1)_1$ the suffix ``$1$'' stands for rotation around $\tau^1$. 
Following Eq.~(\ref{phivac}) we notice that any  rotation around 
$\tau^1$ keeps $\langle\Phi\rangle_v$ invariant. 
Simultaneous $\pi$ rotations in the $U(1)_Y$ group and around an axis along any linear combination of $\tau^3$ and $\tau^2$ keep $\langle\Phi\rangle_v$ invariant, since both the $\pi$ rotations generate sign changes separately.  This defines the  unbroken discrete group $ \mathbb{Z}_2$  and the elements of unbroken group  are defined as
\begin{eqnarray}
H = \l\{\left(1, \,\, e^{\frac{i\alpha}{2}\sigma^1}\right),\, \left(-1, \,\,i\left(c_2 \sigma^2 + c_3\sigma^3\right)e^{i\frac{\alpha}{2}\sigma^1}\right)\r\},
\end{eqnarray}
where $c_2, c_3$ are arbitrary real constants normalized to the unity as $c_2^2 + c_3^2 =1$.
The semi-direct product implies that 
the $U(1)_1$ acts on the ${\mathbb Z}_2$.  
The vacuum manifold is 
\begin{eqnarray}
  \frac{G}{H} = \frac{U(1) \times SO(3)}{O(2)}
\simeq 
 \frac{S^1 \times S^2}{\mathbb{Z}_2}.
\end{eqnarray}

The fundamental group for this symmetry breaking process  
can be calculated as
\begin{eqnarray}
 \pi_1  \left( \frac{G}{H}\right)
\simeq \mathbb{Z}.
\end{eqnarray}
This nontrivial fundamental group indicates  the existence of stable strings. The generator of the unbroken $U(1)$ changes sign as it encircles the string once,  
identifying  this vortex as an Alice string. 

  The scalar and gauge field configurations at a large distance from 
  an Alice string  can be expressed as 
\begin{eqnarray}
\label{Aliceconfiguration}
&& \Phi(R, \theta)  \sim \xi e^{i\frac{\theta}{2}} \left(
\begin{array}{ccc}
 0 &  e^{i\frac{\theta}{2}}     \\
  e^{-i\frac{\theta}{2}} & 0  
\end{array}
\right) 
= \xi e^{i\frac{\theta}{2}} e^{i\frac{\theta}{4}\sigma^3} \sigma^1 e^{-i\frac{\theta}{4}\sigma^3}, \nonumber 
\\
&& A_i \sim - \frac{1}{4 g}\frac{\epsilon_{ij} x^j}{r^2} \sigma^3, \quad 
a_i \sim - \frac{1}{2 e_{}}\frac{\epsilon_{ij} x^j}{r^2}
  \label{eq:Alice-asymptotic}
\end{eqnarray}
with an angular coordinate $\theta$ and the system size $R$. 
 At $\theta =0$ (along the $ x^1$-axis) the scalar field takes its vacuum value
$\Phi(R, \theta=0) = \xi \sigma^1,$
and  the order parameter at any arbitrary $\theta$
 is given by a holonomy action as 
\begin{align}
\Phi(R, \theta) &\sim e^{ie_{} \int {\bf a\cdot dl}} Pe^{i g \int {\bf A\cdot dl}}\, \Phi(R, 0)Pe^{-ig \int {\bf A\cdot dl}} \nonumber\\
 & \sim U_0(\theta) U_3(\theta)\, \Phi(R, 0) U^{-1}_3(\theta),\;
\end{align}
where the holonomies are defined as
\begin{eqnarray}
\label{holonomy}
U_0(\theta) = e^{ie \int_0^\theta {\bf a\cdot dl}} = e^{i\frac{\theta}{2}}\in U(1), \quad U_3(\theta) = Pe^{i g \int_0^\theta {\bf A\cdot dl}} = e^{i\frac{\theta}{4}\sigma^3} \in SU(2).
\end{eqnarray}
These can be understood by using the condition of topological vortex where  the order parameter $\Phi$ is covariantly constant at large distances ($\D_i\Phi \rightarrow 0$ as $R \rightarrow \infty$).  
Since the scalar field configuration of the vortex is space dependent, the unbroken group generators 
also change around the vortex.  In the case of the Alice string as we encircle the string by an angle $\theta$,   the unbroken $U(1)$  generator $Q$ changes as 
\begin{eqnarray}
Q_\theta = U_3(\theta) Q_0 U_3(\theta)^{-1}, \quad U_3(\theta) = e^{\frac{i\theta}{4}\sigma^3},
\end{eqnarray}
where the value of the generator at $\theta=0$ is $Q_0 = \tau^1$. This is true because $Q_\theta$ keeps
Eq.~(\ref{Aliceconfiguration}) invariant. Now it is interesting enough to notice that
 the generator changes its sign after one complete encirclement
 \begin{eqnarray}
Q_{2\pi} = - Q_0.
\end{eqnarray}
This is nothing but the most remarkable feature of the Alice string. 

To find the minimum tension,  the  Bogomol'nyi completion can be performed by considering the critical couplings 
$ \lambda_e = e_{}^2$ and $\lambda_g = g^2$.  
The Bogomol'nyi completion of the tension, that is the static energy per a unit length, is found to be
\begin{eqnarray}
\label{tension}
\mathcal{T} & =& \int d^2x \left[\Tr\left[ F_{12} \pm \frac{g}{2}[\Phi,\Phi^\dagger]\right]^2 
+ \Tr|\D_{\pm} \Phi|^2 
+ \half\left[  f_{12} \pm e_{} \left(\Tr \Phi\Phi^\dagger - 2 \xi^2\right)\right]^2
\pm 2 e_{} f_{12} \xi^2 \right.\Big{]} \nonumber\\ 
 &&\ge 2 e_{}  \xi^2 \left|\int d^2x\; f_{12}\right| = 2 \pi  \xi^2 ,\;
\end{eqnarray}
with $\D_{\pm} \equiv \frac{D_1 \pm i D_2}{2}$.  
The BPS equations are 
\begin{eqnarray}
&& F_{12} \pm \frac{g}{2}[\Phi,\Phi^\dagger]=0, \nonumber\\
&& \D_{\pm} \Phi =0, \nonumber\\
&& f_{12} \pm e_{} \left(\Tr \Phi\Phi^\dagger - 2 \xi^2\right)=0.\label{eq:BPS}
\end{eqnarray}

\begin{figure}[!htb]
\centering
\includegraphics[totalheight=3.5cm]{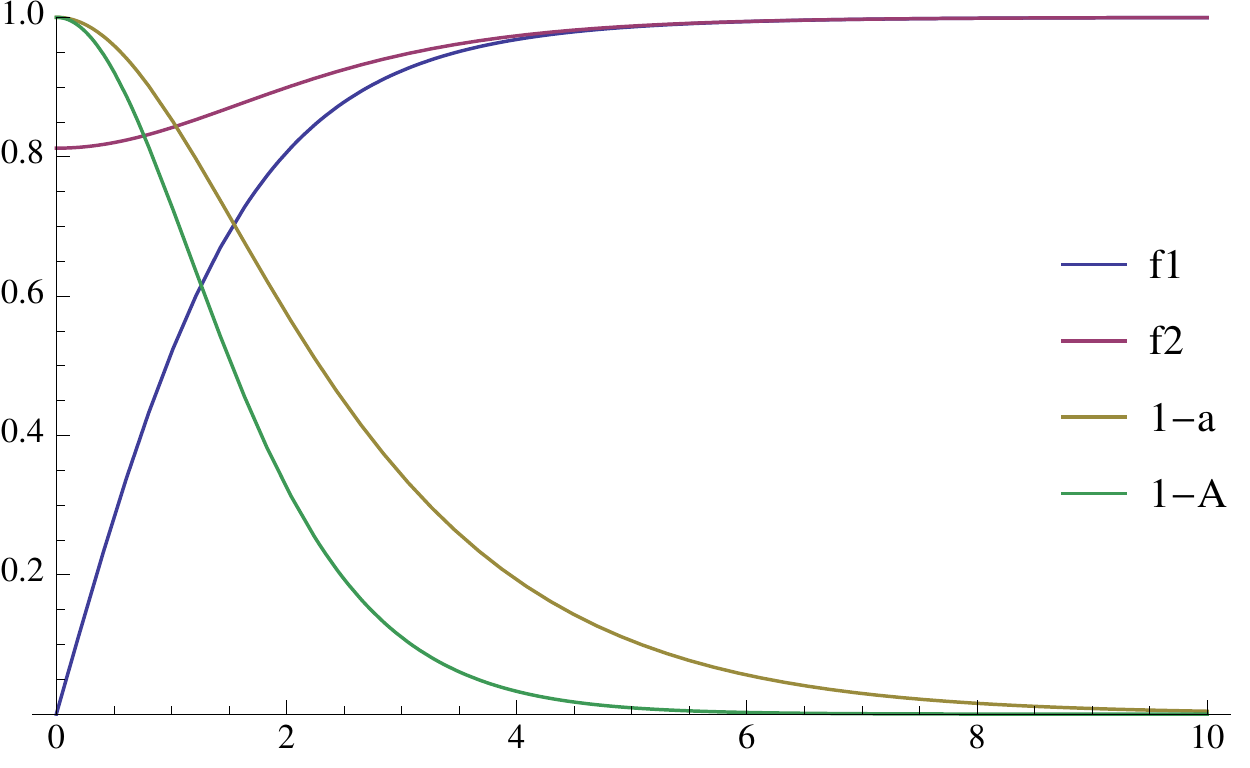}
\caption{The numerical solutions of the profile functions $f_1(r)$ ,  $f_2(r)$, $1 -a(r)$ and $1-A(r)$ are displayed  for a vortex configuration of winding number one as a function of $r$ 
\cite{Chatterjee:2017jsi}. 
\label{profile-Alice}
}
\end{figure}

%%%%%%%%%%%%%%%%%%%%%%%%%%%%%%%%%%%
Now the vortex solutions can be constructed by starting with a vortex ansatz
\begin{eqnarray}\label{phivortex}
&&\Phi(r, \theta) = 
\xi\left(
\begin{array}{ccc}
 0 & f_1(r) e^{i\theta}     \\
f_2(r)   & 0  \end{array}
\right), \nonumber\\
&& A_i(r, \varphi) = -\frac{1}{4g} \frac{\epsilon_{ij} x_j}{r^2}\sigma^3A(r), \quad 
 a_i (r, \theta) = - \frac{1}{2e_{}} \frac{\epsilon_{ij}x_j}{r^2}a(r), 
\end{eqnarray}
where $\{r, \theta \}$ are radial and angular coordinates 
of the two dimensional space, 
respectively. 
The profile functions
$f_1(r), f_2(r), A(r)$ and $a(r)$ depending only on the radial coordinate 
and they 
can be solved numerically 
with the boundary conditions 
$f_1(0) = f_2'(0) = 0,\quad f_1(\infty) = f_2(\infty) =1, A(0)= a(0) = 0, \quad A(\infty)= a(\infty) = 1$.\footnote{
These profile functions eventually satisfy the same equations with 
those for a non-Abelian vortex in $U(N)$ gauge theory 
coupled with $N$ Higgs scalar fields in the fundamental representation 
\cite{Hanany:2003hp,Auzzi:2003fs,Hanany:2004ea,Shifman:2004dr,Eto:2004rz,Gorsky:2004ad,Eto:2005yh,Eto:2006cx,Eto:2006db,Tong:2005un,Eto:2006pg,Shifman:2007ce,Shifman:2009zz}. 
}
The numerical solution for the BPS equation (\ref{eq:BPS})
is displayed in the Fig.~\ref{profile-Alice}.

As can be seen from Eq.~(\ref{eq:Alice-asymptotic}), the vortex solution is generated 
by the $\sigma_3$ generator. This could be $\sigma_2$ or in more general 
$c_2 \sigma_2 + c_3 \sigma_3$ with $c_2^2 + c_3^2=1$. 
Therefore, the Alice string solution is parameterized by a $U(1)$ modulus, 
corresponding to the flux that it carries inside its core. 

The BPS completion helps us to embed our Alice string into 
an ${\cal N}=1$ supersymmetric theory \cite{Chatterjee:2017jsi}, 
in which BPS Alice strings are shown to be 1/2 BPS, preserving a half of supersymmetries.\footnote{While supersymmetric theories with the same matter contents were studied in Ref.~\cite{Davis:1997ny}, Alice properties were not recognized there.
}
In this case, BPS solitons belong to short multiplets of supersymmetry and 
are quantum mechanically stable \cite{Witten:1978mh}.
In this paper, however, the BPS-ness is not necessary.

%%%%%%%%%%%%%%%%%%

\subsection{The Aharonov-Bohm phase around an Alice string}
\subsubsection{The first model: charged doublet}
 Since we have an unbroken $U(1)_1$  symmetry in the bulk, there may exist massless particles in the bulk which are charged under unbroken $U(1)_1$.  These particles may interact with Alice strings which may generate AB phases. To realize this, let us first insert scalar particles with  doublet representation  without the potential.  In this case the doublet scalar field interacts 
	 at low energy only with the unbroken $U(1)$ gauge field in  the bulk since the other $SU(2)$ gauge fields are  massive due to a large VEV of $\Phi$. However, the existence of Alice property generates a nontrivial AB phase due to the unbroken 
 $\mathbb{Z}_2$ defined in Eq.~(\ref{SSB1}).  To confirm this, let us put the fields $\Phi, A_i, a_i$ in its string configurations found in the Alice string analysis, then the doublet field changes around the
vortex as
\begin{eqnarray}
\Psi_\theta &=& e^{i \frac{e_{}}{2} \int_0^\theta a\cdot dl} \l(Pe^{i g \int_0^\theta A\cdot dl} \r)
\left(
\begin{array}{c}
  \psi_1  \\
\psi_2 
\end{array}
\right) 
=
e^{i \frac{\theta}{4}} e^{i \frac{\theta}{4}\sigma^3}
 \left(
\begin{array}{c}
  \psi_1  \\
\psi_2 
\end{array}
\right) 
=  \left(
\begin{array}{c}
e^{i \frac{\theta}{2}}  \psi_1  \\
\psi_2 
\end{array}
\right) .
\end{eqnarray}

As it can be noticed that after a complete encirclement, the doublet field gets an AB phase as
\begin{eqnarray}
\label{thetadoublet1}
\Psi_{2\pi} &=& e^{i \frac{e_{}}{2} \oint a\cdot dl} \l(Pe^{i g \oint A\cdot dl} \r)
\left(
\begin{array}{c}
  \psi_1  \\
\psi_2 
\end{array}
\right) = \left(
\begin{array}{c}
-  \psi_1  \\
\psi_2 
\end{array}
\right).
\end{eqnarray}
This has interesting physical consequences.  We may define charges by the eigenvalues of the $U(1)_1$ generator as $Q = \sigma^1$, since it is the only gauge symmetry group which lives at the bulk or far away from the vortex core.
So we describe the system with eigen states  of charge operator as
\begin{eqnarray}
|+\rangle = \frac{1}{\sqrt 2}\left(
\begin{array}{c}
1  \\
1
\end{array}
\right),\qquad |-\rangle = \frac{1}{\sqrt 2}\left(
\begin{array}{c}
- 1  \\
1
\end{array}
\right).
\end{eqnarray}
According to Eq.~(\ref{thetadoublet1}), when the charged state $|+\rangle$ encircles around the vortex it becomes 
$|- \rangle$.  
Therefore, the charge conjugation symmetry 
cannot be defined globally.

\subsubsection{The second model: chargeless doublet}

 Since we have chargeless doublet scalar particles in the bulk, these particles realize an AB phase only from the non-Abelian $SU(2)$ gauge field configuration of the Alice string.  The doublet field changes around the
vortex as  
\begin{eqnarray}
&& 
\Psi_\theta =  \l(Pe^{i g \int_0^\theta A\cdot dl} \r)
\left(
\begin{array}{c}
  \psi_1  \\
\psi_2 
\end{array}
\right) 
= e^{i \frac{\theta}{4}\sigma^3}
 \left(
\begin{array}{c}
  \psi_1  \\
\psi_2 
\end{array}
\right) 
=  \left(
\begin{array}{c}
e^{i \frac{\theta}{4}}  \psi_1  \\
e^{-i \frac{\theta}{4}}\psi_2 
\end{array}
\right).
\end{eqnarray}
As it can be noticed that after a complete encirclement the doublet field receive an AB phase as
\begin{eqnarray}
\label{thetadoublet}
\Psi_{2\pi} &=&  Pe^{i g \oint A\cdot dl} 
\left(
\begin{array}{c}
  \psi_1  \\
\psi_2 
\end{array}
\right) = i \left(
\begin{array}{c}
  \psi_1  \\
 - \psi_2 
\end{array}
\right).  \label{eq:AB2}
\end{eqnarray}
According to Eq.~(\ref{thetadoublet}), when the charge state $|+\rangle$ encircles around the vortex, it becomes 
$-i |-\rangle$.  In this case, the positive charge not only transforms into a negative one, but it also acquires an over all $\pi$ phase.

%%%%%%%%%%%%%%%%%%%%%%%%%%%%%%%%%%%%%%%%%%%%%%%%
\section{Aharonov-Bohm defects in the first model with a charged doublet}\label{sec:model1}

So far we have discussed the AB phase of the doublet field
around an Alice string. Here and in the next section, 
we answer to the question what happens when such a field with a non-trivial AB phase acquires a VEV.
In this section, we discuss it in the first model 
in the presence of a charged doublet scalar field with $e_{\psi} = \half$.

\subsection{The charged doublet potential and symmetry breaking}
We now switch on the full potential in Eq.~(\ref{potential}), including the interaction 
$V_{\rm int}^{(1)}(\Phi, \Psi)$. 
The potential generates a nonzero VEV for the doublet field, $\langle\Psi\rangle \ne 0$ if we set a negative bare mass term, $M^2 < 0$. 
As we see below, 
the doublet field breaks the unbroken symmetry group $O(2)$ completely 
or into a $\mathbb{Z}_2$ subgroup, depending on the parameter choice.   
To understand the vacuum
in the presence of the large hierarchy
$\langle |\Phi|\rangle \gg \langle |\Psi|\rangle $, 
we set the triplet field in its 
vacuum value as $\Phi = \xi \sigma^1$ and substitute it into the potential in Eq.~(\ref{potential}):  
\begin{eqnarray}
 V(\Phi, \Psi) &=& \l[ M^2 +  \xi^2( 2 \lambda_1   + \lambda_2)\r] \Psi^\dagger\Psi + \lambda_\psi \l(\Psi^\dagger\Psi\r)^2 + \xi \mu^{(1)}  \l( \Psi_c^\dagger \sigma^1 \Psi +   \Psi^\dagger \sigma^1 \Psi_c\r) .
\end{eqnarray}
We assume the condition $- M^2 > 2 \xi^2( \lambda_1   + \lambda_2)$ to trigger the symmetry breaking.

In the case of $\mu^{(1)}  = 0$, the potential has 
an enlarged $SO(4)$ symmetry and the vacuum manifold is $S^3$, 
defined by 
\begin{eqnarray}
\label{v2}
|\psi_1|^2 + |\psi_2|^2 =  - \frac{ M^2 +  \xi^2(2 \lambda_1   + \lambda_2)}{2 \lambda_\psi} = v_2^2,
\end{eqnarray}
where we have inserted 
$\Psi = \left(
\begin{array}{c}
\psi_1  \\
\psi_2
\end{array}
\right)$.

In the case of $\mu^{(1)}  \ne 0$, we insert
$
\Psi_c = \left(
\begin{array}{c}
\psi_2^*  \\
- \psi_1^*
\end{array}
\right), \psi_1 = \psi_{11} + i  \psi_{12}, \psi_2 = \psi_{21} + i  \psi_{22}
$
into the potential to find
\begin{eqnarray}
\label{potpsi}
V(\Phi, \Psi) 
&=& - \l(2 \lambda_\psi  v_2^2 + 2 \xi \mu^{(1)} \r) \l(  \psi_{11}^2  +   \psi_{22}^2\r) 
 +2  \l(  \xi \mu^{(1)}  -  \lambda_\psi  v_2^2   \r) \l(   \psi_{12}^2 +  \psi_{21}^2\r)  \nonumber\\ 
 && + \lambda_\psi \l( \psi_{11}^2  +   \psi_{22}^2 +  \psi_{12}^2 +  \psi_{21}^2 \r)^2.
\end{eqnarray}
The symmetry of this potential is reduced from $SO(4)$ to 
$U(1) \times U(1)$. 
We now discuss the following two cases separately depending on the sign of the second term:
the case I ($ \lambda_\psi  v_2^2 <   \xi \mu^{(1)} $) 
and the case II  ($ \lambda_\psi  v_2^2 >   \xi \mu^{(1)} $).

The case I: 
the vacuum manifold is $S^1$ described by
\begin{eqnarray}
 \psi_{11}^2  +   \psi_{22}^2 = \frac{\l( \lambda_\psi  v_2^2 +  \xi \mu^{(1)} \r)}{ \lambda_\psi  }= \xi_{\psi_1}^2,\qquad   \psi_{12} = \psi_{21} = 0.  \label{eq:vac1}
\end{eqnarray}
The fact that the potential is invariant under $S^1 \times S^1$ 
can be understood as follows.
The actual $SO(4)$ symmetry in the $\mu^{(1)} =0$ case is realized by taking into account the full symmetry group as $SU(2)_L \times SU(2)_R$.
In the case of $\mu^{(1)}  \ne 0$, the symmetry transformations 
$U_L = e^{i \frac{\alpha}{2}\sigma^1}\in SO(2) \subset SU(2)_L, \quad
e^{i \frac{\beta}{2}\sigma^1}\in SU(2)_R$ act on 
$(\Psi, \Psi_c)$
as
\begin{eqnarray}
 (\Psi, \Psi_c) &&\to U_L(\alpha) (\Psi, \Psi_c) U_R(\beta)\nonumber \\ 
&& =  \left(
\begin{array}{ccc}
 \cos\frac\alpha 2 &   i\sin\frac \alpha 2  \\
 i\sin\frac \alpha 2 &   \cos\frac \alpha 2  
\end{array}
\right)\left(
\begin{array}{cc}
\psi_1& {\psi_2*} \\
 {\psi_2} & {-\psi_1^*}
\end{array}
\right)
\left(
\begin{array}{ccc}
 \cos\frac\beta 2 &   i\sin\frac \beta 2  \\
 i\sin\frac \beta 2 &   \cos\frac \beta 2  
\end{array}
\right)
\end{eqnarray}
In terms of the components, they are
\begin{eqnarray}
\label{psi''}
&&\psi_{11}' = \cos \frac{\alpha-\beta }2 \psi_{11} + \sin\frac {\alpha-\beta } 2 \psi_{22},\nonumber\\
 &&\,\, \psi_{12}' = \cos\frac{\alpha+\beta } 2 \psi_{12} + \sin\frac{\alpha+\beta } 2 \psi_{21},\nonumber \\
&& \psi_{22}' = \cos\frac{\alpha-\beta } 2 \psi_{22} - \sin\frac{\alpha-\beta } 2 \psi_{11},\nonumber\\ &&\,\, \psi_{21}' = \cos\frac{\alpha+\beta } 2 \psi_{21} - \sin\frac {\alpha+\beta } 2 \psi_{12}.  
\end{eqnarray}
Here $U_L(\alpha)$ is a gauge transformation whereas $U_R(\beta)$ is a global symmetry transformation.
We have two circles parametrized by the two parameters $\frac {\alpha-\beta } 2$ and $\frac {\alpha+\beta } 2$. In the vacuum where
$\psi_{12} = \psi_{21} = 0 $, we have an unbroken global $U(1)$, 
if we choose a gauge where $\alpha = \beta$.
 We denote the unbroken $U(1)$ group as color-flavor locked  $U(1)_{L+R}$ parametrized by $\alpha$.

 In the case II, the potential in Eq.~(\ref{potpsi}) with $ \lambda_\psi  v_2^2 >  \xi \mu^{(1)} $ 
 yields the vacua,\footnote{
In addition to the true vacua, the Hamiltonian has two more critical points:
$\psi_{12} = \psi_{21} = 0,  \; \psi_{11} = \psi_{22} = 0$ and 
$\psi_{12}^2  +   \psi_{21}^2 = 
( \lambda_\psi  v_2^2 -  \xi \mu^{(1)} )/\lambda_\psi,\;   \psi_{11} = \psi_{22} = 0$.
It can be checked by computing the determinant of a Hessian matrix that 
the first is a local maximum and the second gives saddle points.
} 
 \begin{eqnarray}
&& \psi_{11}^2  +   \psi_{22}^2 = \frac{\l( \lambda_\psi  v_2^2 +  \xi \mu^{(1)}  \r)}{ \lambda_\psi  }= \xi_{\psi}^2,\qquad   \psi_{12} = \psi_{21} = 0.
\end{eqnarray}
 These vacua $S^1$ are the same as Eq.~(\ref{eq:vac1}) in the case I.   

For both the cases, the vacua are $S^1$
  parametrized by $\frac{\alpha-\beta }2$. 
 If we set $ \alpha = -\beta = \gamma$ as a gauge choice, 
the generic vacua are found to be
\begin{eqnarray}
\label{dwvacuum}
\Psi_\gamma = \xi_\psi \left(
\begin{array}{c}
 \cos\gamma  \\
- i \sin\gamma
\end{array}
\right).
\end{eqnarray}
 We set our vacuum at $\gamma =0$ in the followings for simplicity.

 To calculate VEVs of the triplet and doublet in the case of $\mu^{(1)} = 0$,  we may write the potential in terms of both the VEVs as
\begin{eqnarray}
&&V(|\Phi|_v, |\Psi|_v) =  |\Phi|_v^4\lambda_e\l(1 - \frac{\xi^2}{|\Phi|_v^2}\r)^2 - m^2 |\Psi|_v^2 
 + \lambda_\psi |\Psi|_v^4  + (2 \lambda_1 
 +  \lambda_2) |\Psi|_v^2  |\Phi|_v^2.
\end{eqnarray}
 Here we have defined $M^2 = - m^2$. By minimizing this potential, we find the solution is 
 \begin{eqnarray}
 \label{backreactionvev}
|\Phi|_v 
= \sqrt{\frac{4 \lambda_e\lambda_\psi \xi^2- (2 \lambda_1 
 +  \lambda_2)m^2}{4 \lambda_e\lambda_\psi - (2 \lambda_1 
 +  \lambda_2)^2}}, \quad 
 |\Psi|_v 
 = \sqrt{\frac{2 \lambda_e\l[m^2 - \xi^2(2 \lambda_1 
 +  \lambda_2)\r]}{4 \lambda_e\lambda_\psi - (2 \lambda_1 
 +  \lambda_2)^2}}.
\end{eqnarray}
 %%%%%%%%%%%%%%%%%%%%%%%%%%%%%%%%%%%%%%%%%%%%%%%%%%%%%%%%%%
 \subsection{The non-interactive case: a global vortex with a fractional flux}
In the last subsection, we have discussed the symmetries of the potential in 
the case in which the triplet is set to its vacuum value $\Phi_v = \xi\sigma^1$ 
and have studied the vacuum of the doublet in this case. Now we are going to ask the question what happens if we set the triplet field configuration to 
an Alice string. To understand the situation, we first consider $\xi \gg \xi_\psi$ to remove backreaction. In this case, the Alice string behaves as a background and  
we study a nonzero VEV $\Psi_v$ of the doublet field in this background. 
 We thus substitute the large distance Alice string configurations in 
  Eq.~(\ref{Aliceconfiguration}) into the action.

However, as we have discussed, 
the doublet field gets an AB phase 
in the presence of an Alice string. 
Then, the VEV $\Psi_v$ of the doublet would be 
non-single-valued after one encirclement of the Alice string, 
which is inconsistent. 
To overcome this problem,  
when  $\mu^{(1)} =0$,
the system would choose an energetically favorable  configuration,  
in which the AB phase of the doublet should be nullified  by 
a global transformation for the doublet. 
So the doublet does not get any winding even if it takes a part into 
the formation of a vortex. 
More precisely, 
the first component of the doublet field ($|\psi_1|_v$) receives the winding  
which 
arises due to the AB phase, but it should be canceled by a simultaneous global rotation. 
Let us imagine the following two steps (although they occur simultaneously in practice). In first step, we take the doublet at $\theta=0$ as
$
\Psi|_{\theta=0} = \frac{|\Psi|_v}{\sqrt{2}} \left(
\begin{array}{ccc}
1   \\
 1
\end{array}
\right).
$
If we go along a circle encircling the Alice string, it would acquire 
the AB phase to become 
$\Psi_v(\theta)=  \frac{|\Psi|_v}{\sqrt{2}} \left(
\begin{array}{ccc}
e^{i\frac{\theta}{2}}   \\
 1
\end{array}
\right)$, which is not single-valued.
In the second stage,
 a global rotation is created to make the doublet single-valued (constant).

In order to construct a full numerical vortex solution, we may take into account the backreaction due to the interaction between doublet and triplet field. 
In this case, we use VEVs of the scalar fields in Eq.~(\ref{backreactionvev}).
We consider a vortex ansatz
\begin{eqnarray}
\label{phivortex2}
&&\Phi(r, \theta) = 
|\Phi|_v \left(
\begin{array}{ccc}
 0 & f_1(r) e^{i\theta}     \\
f_2(r)   & 0  \end{array}
\right),   
\quad \Psi_v = |\Psi|_v \left(
\begin{array}{c}
 \psi(r) \\
0
\end{array}
\right),
\nonumber\\
&&
A_i(r, \theta) = -\frac{\c}{4g} \frac{\epsilon_{ij}x_j}{r^2}\sigma^3A(r),\quad
a_i (r, \theta) = - \frac{\c}{2e_{}} \frac{\epsilon_{ij}x_j}{r^2}a(r), 
\end{eqnarray}
where $\{r, \theta \}$ are radial and angular coordinates 
of the two dimensional space,
respectively. 
The profile functions
$f_1(r), f_2(r), A(r), a(r)$ and $\psi(r)$ depend only on the radial coordinate
and they can be solved numerically with the boundary conditions
$f_1(0) = f_2'(0) = \psi(0) = 0,\quad f_1(\infty) = f_2(\infty) = \psi(\infty) = 1, A(0)= a(0) = 0, \quad A(\infty)= a(\infty) = 1$.
$\c$ is the fractional flux arising due to backreaction, which 
becomes $\c \simeq 1$ when $\xi \gg v_2$. 
To understand the effect of backreaction  on fluxes, we insert the large distant configurations 
of the scalar fields into the  Hamiltonian while we keep the gauge fields without fixing any special form. We insert $A_i = A_i^3 \tau^3$  and the gradient  terms of scalar fields, to give
\begin{eqnarray}
\mathcal{H}_{\rm grad} &&\sim |\Phi|_v^2\l(\p_i\theta - (e_{} a_i + g A_i^3)\r)^2 +  |\Phi|_v^2\l(e_{} a_i - g A_i^3\r)^2
+ \frac{ |\Psi|_v^2}{4}\l(e_{} a_i + g A_i^3\r)^2.
\end{eqnarray}
After minimizing we find the solution
\begin{eqnarray}
e_{} a_i + g A_i^3 = \frac{|\Phi|_v^2\p_i\theta}{ |\Phi|_v^2 +  \frac{ |\Psi|_v^2}{4}}, \qquad e_{} a_i = g A_i^3.
\end{eqnarray}
We thus find
\begin{eqnarray}
\c = \frac{|\Phi|_v^2}{ |\Phi|_v^2 + \frac{ |\Psi|_v^2}{4}}.
\end{eqnarray}
This makes the Abelian and non-Abelian fluxes fractional as
\begin{eqnarray}
\oint a \cdot dl = \frac{  \pi |\Phi|_v^2}{e_{}( |\Phi|_v^2 + \frac{ |\Psi|_v^2}{4})},\qquad
\oint A^3 \cdot dl = \frac{  \pi |\Phi|_v^2}{g( |\Phi|_v^2 + \frac{ |\Psi|_v^2}{4})}.
\end{eqnarray}
We thus have found that 
the fractional nature of fluxes is 
due to the existence of the doublet VEV.

\begin{figure}[!htb]
\begin{tabular}{ccc}
\includegraphics[totalheight=3.0cm]{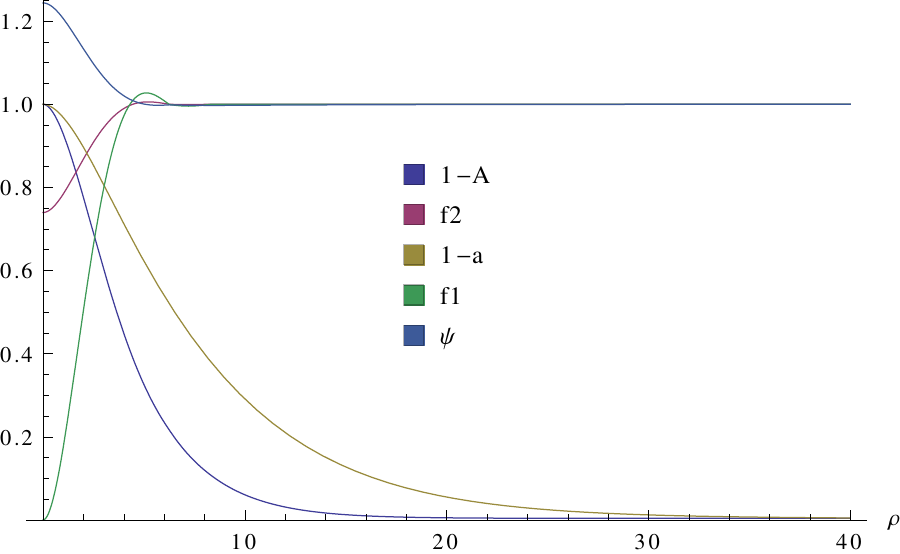} &
\includegraphics[totalheight=3.0cm]{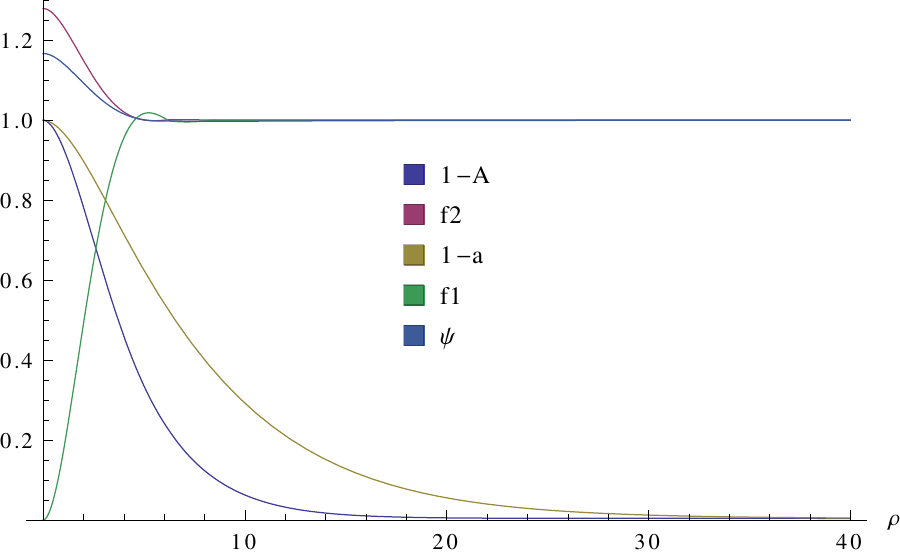} &
\includegraphics[totalheight=2.5cm]{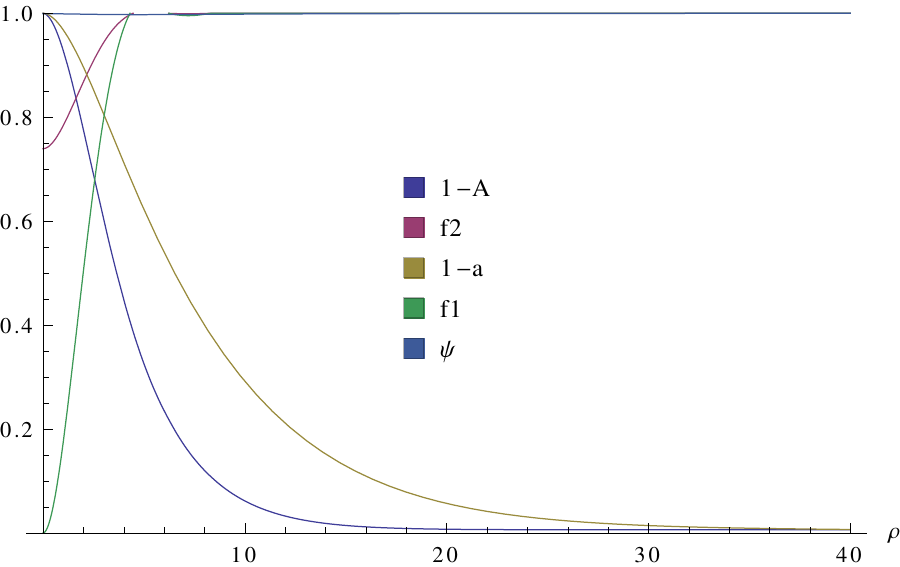}\\
(a) & (b) & (c)
\end{tabular}
\caption{Profile functions of a global vortex with a fractional flux. 
The parameters choice is: 
(a) $\lambda_g = 1, \lambda_1 = 0.03, \lambda_2 = 0.03, \lambda_e = 
1, \lambda_\psi = 5, m = 1, \xi = 2, e_{} = 0.05, g = 0.1$.
(b) $\lambda_g = 0.2$, and the rests are the same with (a). 
(c) 
$\lambda_g =1, \lambda_1 = \lambda_2 =0$
and the rests are the same as (a) and (b). 
One may notice that $\psi(r)$ gets a trivial shape if $\lambda_1 = \lambda_2 = 0$.
}
\label{profileGV}
\end{figure}

Now we write the equations of motion of all profile functions as
\begin{eqnarray}
&&-  r\frac{\p}{\p r} \l(\frac{A'}{r}\r) + 4 g^2 |\Phi|_v^2  
\l[\l( \frac{a + A}{2} - \frac{1}{\c}\r)f_1^2+ \frac{(A - a)}{2 }f_2^2\r] 
+  g^2|\Psi|_v^2 (a + A)\psi^2 = 0, \nonumber\\
&&-  r\frac{\p}{\p r} \l(\frac{a'}{r}\r) + 4 e_{}^2|\Phi|_v^2  
 \l[\l( \frac{a + A}{2} - \frac{1}{\c}\r)f_1^2+ \frac{( a - A)}{2 }f_2^2\r] 
 +  e_{}^2|\Psi|_v^2 (a + A)\psi^2 = 0, \nonumber\\
&& - \frac{1}{r}\frac{\p}{\p r}(r f_1' ) +\l(1 - \c^2\frac{a + A}{2}\r)^2 \frac{f_1}{ r^2}  
  + |\Phi|_v^2\l[\lambda_g \l(f_1^2 - f_2^2\r)f_1
  + \frac{\lambda_e}2\l(f_1^2 + f_2^2 - 2\frac{\xi^2}{|\Phi|_v^2}\r)f_1\r] \nonumber\\ 
&&\quad\quad +  (\lambda_1 + \lambda_2) |\Psi|_v^2 \psi^2 f_1 = 0, \nonumber\\ 
 && - \frac{1}{r}\frac{\p}{\p r}(r f_2' ) 
 +  \c^2\frac{(a - A)^2}{4 r^2}f_2 + |\Phi|_v^2\l[   \lambda_g\l(  f_2^2 - f_1^2\r) f_2
+  \frac{\lambda_e}2\l(f_1^2 + f_2^2 - 2\frac{\xi^2}{|\Phi|_v^2}\r)f_2\r] \nonumber\\
&& \quad\quad + \lambda_1 |\Psi|_v^2 \psi^2  f_2 = 0, \nonumber\\ 
 &&-  \frac{1}{r}\frac{\p}{\p r}(r \psi' ) +  \frac{\c^2}{8 r^2}(a + A)^2\psi 
   +\l[ m^2   + 2 \lambda_\psi |\Psi|_v^2 \psi^2 + \lambda_1 (f_1^2 + f_2^2) |\Phi|_v^2
  +  \lambda_2  f_1^2  |\Phi|_v^2 \r] \psi = 0 .\nonumber\\
\end{eqnarray}
Numerical solutions of these equations are portrayed in Fig.~\ref{profileGV}. 

%%%%%%%%%%%%%%%%%%%%%%%%%%%%%%%%%%%%%%%%%%%%%%%%%%%%%%%%%%%
 \subsection{The interactive case: an Alice string confined by a soliton}
 So far we discussed AB phases, symmetry breaking pattern and 
 a global vortex due to a nonzero VEV of the doublet 
 in the case of $\mu^{(1)} =0$ (in the absence of the interaction depending 
 on the relative phase between the triplet and doublet). 
 In this subsection, we switch on $\mu^{(1)}$ 
 and discuss what happens for the global vortex discussed in the last subsection.

 Let us first discuss the  case  $\xi \gg \xi_\psi$ to remove 
 a backreaction of the doublet VEV to the triplet, 
 and later we discuss the general case.
  In  this case, the Alice string is heavy and 
  behaves as a background. 
 So we set the large distance configurations defined in 
 Eq.~(\ref{Aliceconfiguration}) 
 as our background configuration.
In the presence of an Alice string, 
the doublet field receives an AB phase which 
makes the doublet VEV non-single-valued after one encirclement of the Alice string as discussed in the previous subsection.  
In the case of $\mu^{(1)}  \ne 0$, 
here  we propose that the system would choose 
an energetically favorable configuration which generates 
a domain wall or soliton to preserve single-valuedness of the doublet VEV.  
It can be understood clearly by introducing an additional 
phase $\phi(x)$ of the doublet 
to ensure single-valuedness of the doublet field. $\phi(x)$ changes as the doublet encircles the Alice string together with the AB phase. 
We thus have an ansatz for the doublet as
\begin{eqnarray}
\Psi_{\rm DW} \sim \xi_\psi \left(
\begin{array}{c}
 e^{i \l(\frac{\theta + \phi(\theta)}2  \r)}  \\
0
\end{array}
\right),\,\, \phi(0) = 0,\,\, \phi(2\pi) = -2\pi ,\label{eq:doublet-ansatz1}
\end{eqnarray}
where $\phi(\theta)$ is a decreasing function and 
the boundary condition 
keeps the doublet single-valued 
(constant).
We substitute this ansatz 
 to the  Hamiltonian density in the presence of the Alice string configuration 
 in Eq.~(\ref{eq:Alice-asymptotic})
 at large distances, 
where the potentials are given in Eqs.~(\ref{eq:VPsi}) and (\ref{eq:int1}). 
We thus obtain the effective Hamiltonian of the doublet
\begin{eqnarray}
\mathcal{H}_{\rm DW}/\xi_{\psi}^2 \sim \frac{1}{4}\l[(\p_i\phi)^2 + 8 \mu_\psi^2 \l(1 - \cos\phi\r)\r], 
\quad \mu_\psi^2 = \xi\mu,
\end{eqnarray}
which is nothing but 
the sine-Gordon model. 
With the boundary condition for $\phi$ 
in Eq.~(\ref{eq:doublet-ansatz1}),
we inevitably encounter a single kink 
$
\phi(x) = 4 \tan^{-1}e^{\pm 2 \mu_m x}$,
with the energy density per the unit area, given as $T_{\rm DW} = 4 \mu_\psi\xi_{\psi}^2$.

\begin{figure}[htbp]
\centering
\begin{tabular}{ccc}
\includegraphics[totalheight=3.5cm]{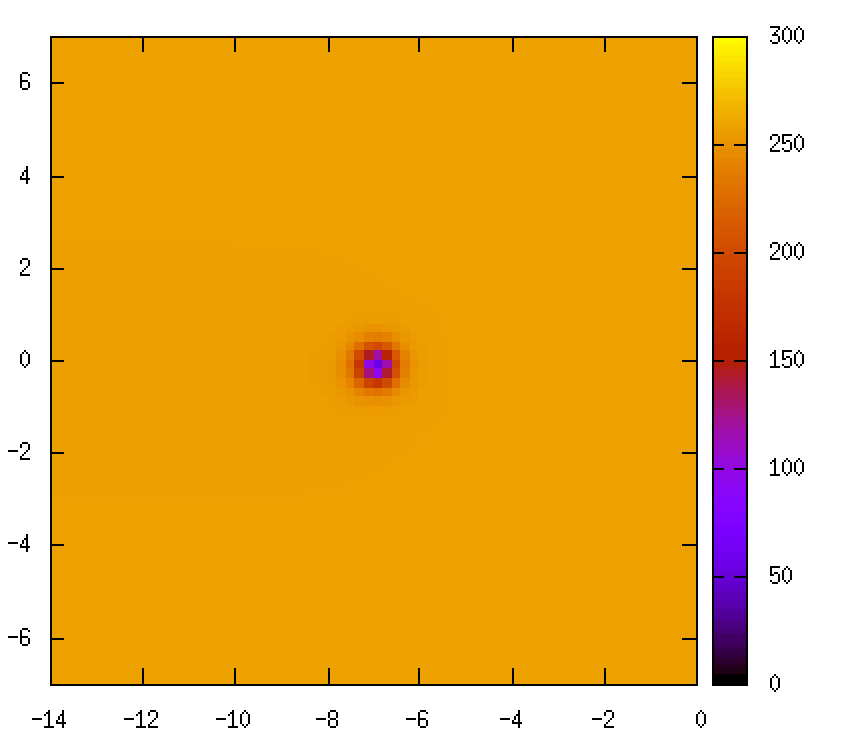} &
\includegraphics[totalheight=3.5cm]{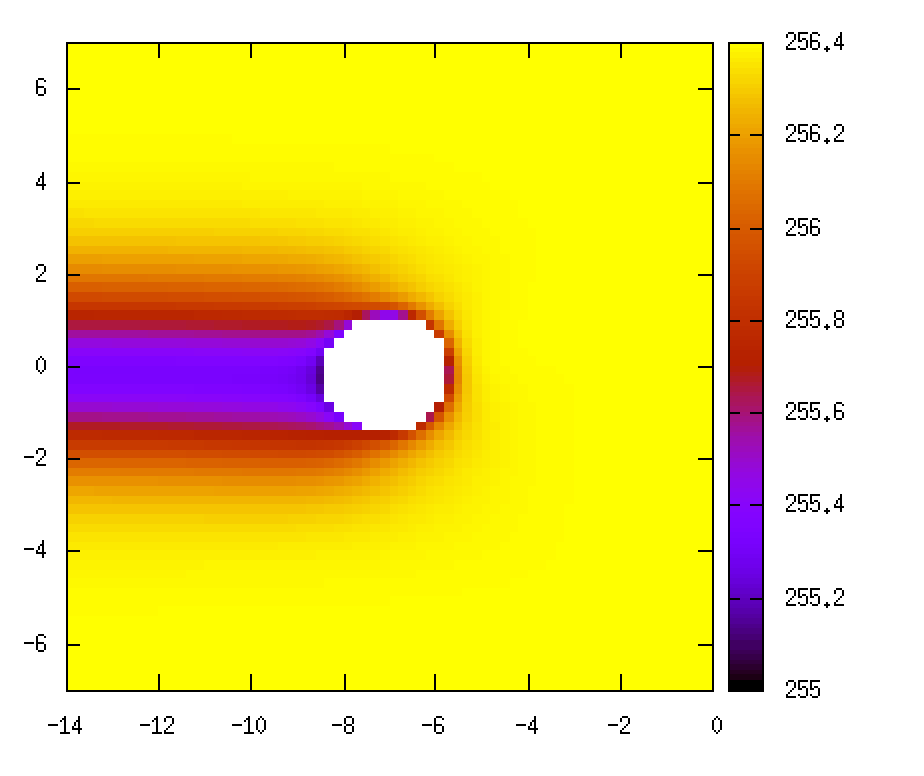} &
\includegraphics[totalheight=3.5cm]{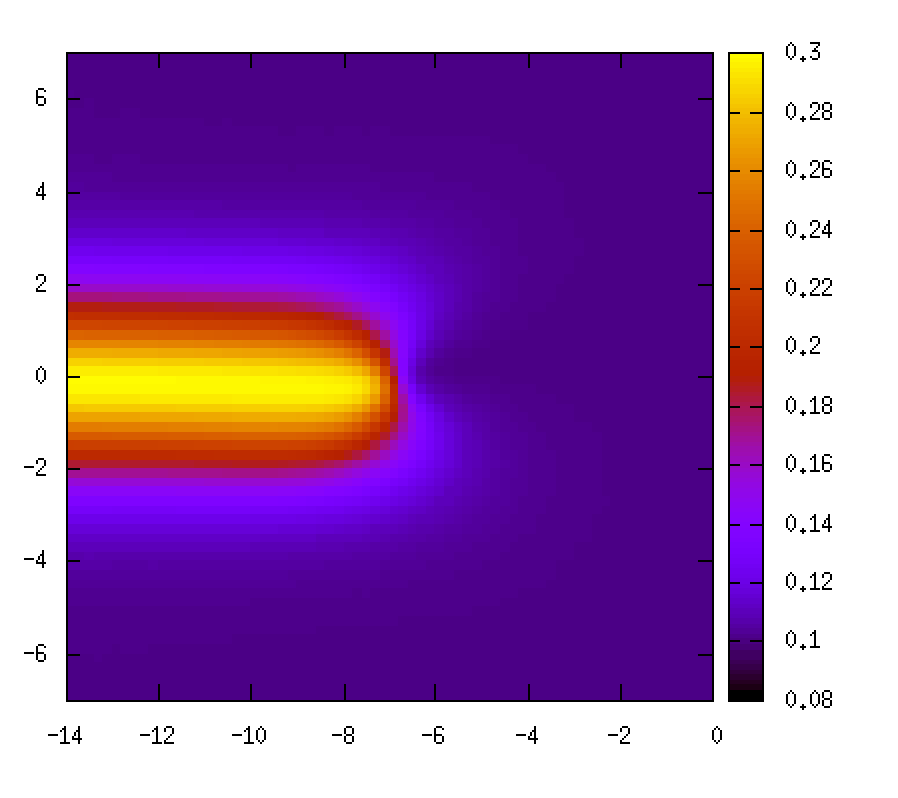}  \\
(a) & (b) & (c) \\
\includegraphics[totalheight=3.5cm]{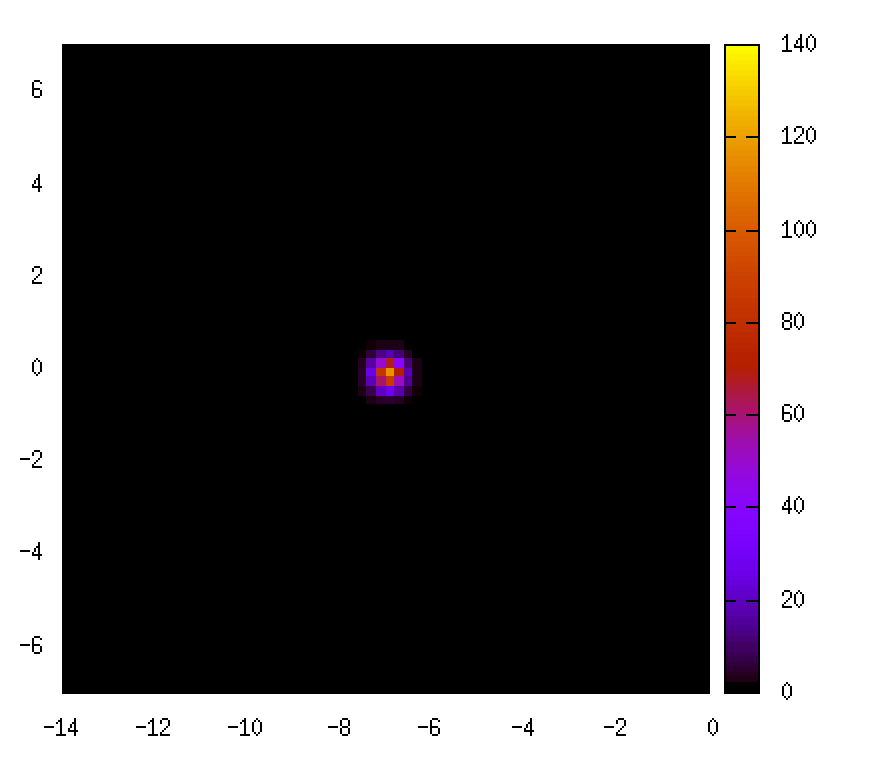} &
\includegraphics[totalheight=3.5cm]{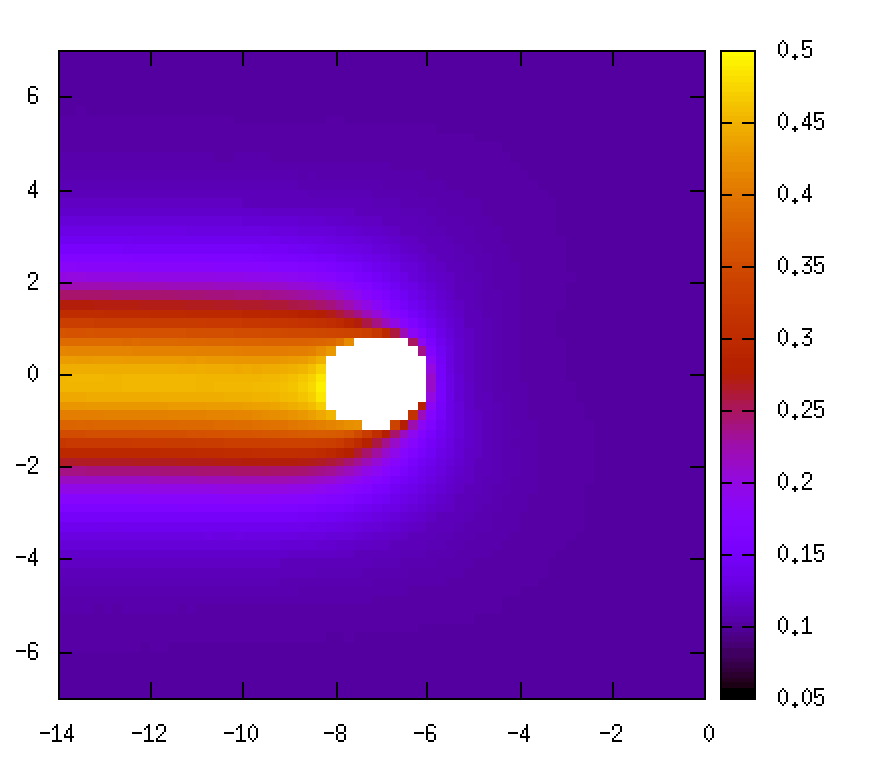} &
\includegraphics[totalheight=3.5cm]{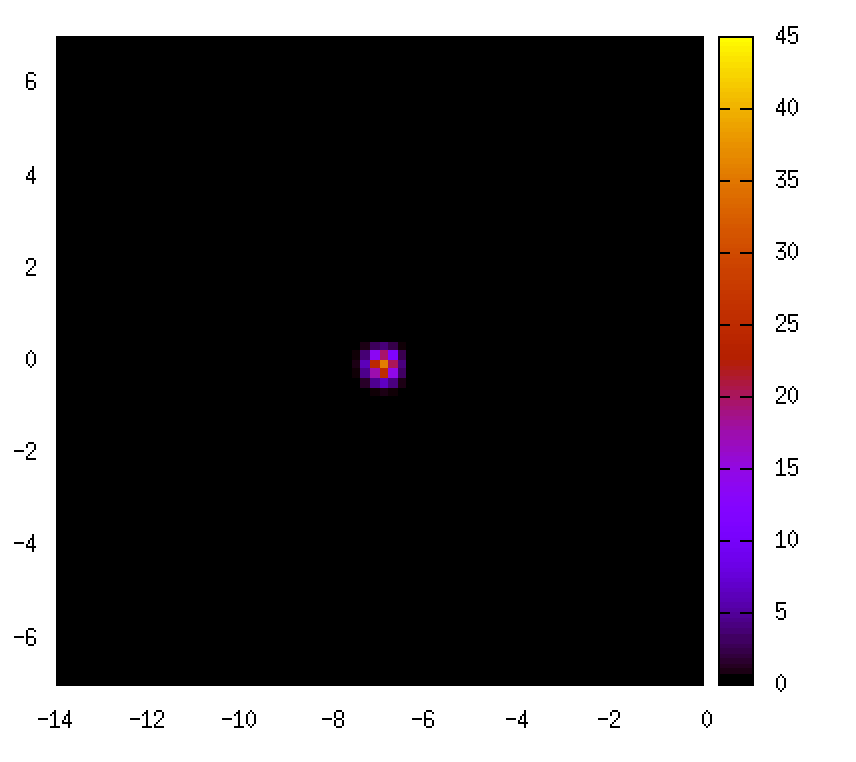}\\
(d) & (e) & (f) 
\end{tabular}
\caption{A full 2D numerical simulation for an Alice string confined by a soliton in the first model with a charged doublet. 
    The all subfigures are plots for different gauge invariants:
(a) the gauge invariant 
$4 \Tr\Phi^2 \Tr{\Phi^\dagger}^2$ which 
is zero at the center of the Alice string; 
If we insert our Alice string ansatz into it, it is $\sim \xi^4 f(r)^2g(r)^2$.
(b) the same plot with (a) in a smaller range  $256.4 - 255$
to clearly show the existence of a domain wall. 
(The white circular region at the center lies outside the range.)
 (c) the interaction potential $V_{\rm int}^{(1)} + \frac{2\xi m^2}{\lambda_\psi}$, 
 in which the domain wall is clearly visible.
 (d) the energy density $\mathcal{H}(x,y)$, 
 in which the domain wall is not visible clearly because of very low intensity. 
  (e) the same plot with (d) in a smaller range $0.05 - 0.5$ to show the domain wall.
  (f) the flux squared $\Tr F_{12}^2 =\sum_{a=1}^3{F_{12}^a}^2$,
 in which the confined magnetic field can be seen clearly. 
 The parameter choices are  $\xi = 0.1,   e_{} = 0.5,  g = 1,
     \mu^{(1)}  = 0.1 , \lambda_\psi= 1, m =  0.1, \lambda_e= 2, \lambda_g=1 $. 
     }
\label{DW_Alice1}
\end{figure}

 To confirm our claim, we solve the full two-dimensional 
 equations of motion numerically by the relaxation method, 
 as shown in Fig.~\ref{DW_Alice1}.
 To do this computation, we have used a $500\times500$ square lattice with a lattice spacing $0.2$.  
The details of a numerical method can be found in Appendix \ref{sec:appendix}. 
   Since the configuration is unstable in the sense that the wall pulls the Alice string to infinity, 
   this configuration is a snapshot after the shape is converged.
   
It may be interesting to emphasize that 
the spatial rotation around the string is spontaneously broken once the interaction term proportional to 
$\mu^{(1)}$ is introduced.

 %%%%%%%%%%%%%%%%%%%%%%%%%%%%%%%%%%%%%%%%%
%%%%%%%%%%%%%%%%%%%%%%%%%%%%%%%%%%%%%%%%%%%%%%%%%

\section{Aharonov-Bohm defects in the second model with a chargeless doublet}\label{sec:model2}
In this section, we consider the second model with a chargeless doublet 
with the potential term in Eq.~(\ref{eq:int2}).

\subsection{The chargeless doublet potential and symmetry breaking}
Here we study the vacuum of the doublet field with keeping the triplet in its original vacuum, 
when the triplet VEV is much larger than the doublet VEV. 
Following Eq.~(\ref{v2}) and  inserting the vacuum configuration of the triplet ($\Phi = \xi \sigma^1$) into the potential we  find the full potential as
\begin{eqnarray}
V(\Phi, \Psi)
&=&- v_2^2 2 \lambda_\psi \l(|\psi_1|^2+ |\psi_2|^2\r) +  \lambda_\psi\l(|\psi_1|^2+ |\psi_2|^2\r)^2 + \xi^2 \mu^{(2)} \l|\psi_1^2 - \psi_2^2\r|^2
\end{eqnarray}
with $ v_2^2 = \frac{    - M^2 +  \xi^2(2 \lambda_1   + \lambda_2)}{2 \lambda_\psi} $. The vacuum manifold is found to be
\begin{eqnarray}
|\psi_1| = |\psi_2| = \pm \frac{v_2}{\sqrt 2}, 
\quad \text{or}\quad \Psi_v = \frac{v_2 e^{i \beta}}{\sqrt 2}
\left(
\begin{array}{ccc}
  1  \\
 \pm 1   
\end{array}
\right).
\end{eqnarray}
At the vacua the fields $\psi_1$ and $\psi_2$ are in the same phase or 
have a $\pi$ phase difference. 
The doublet VEV breaks $O(2)$ completely including $\mathbb{Z}_2 \in O(2)$.

We thus can expect a domain wall configuration interpolating between $|+\rangle$ and $|-\rangle$, 
and we call this as an ``Alice wall.'' 
Different from the first model, this model admits a topologically stable domain wall.

%%%%
\subsection{The non-interactive case: a global vortex with a fractional flux}
\begin{figure}[!htb]
\centering
\includegraphics[totalheight=3.5cm]{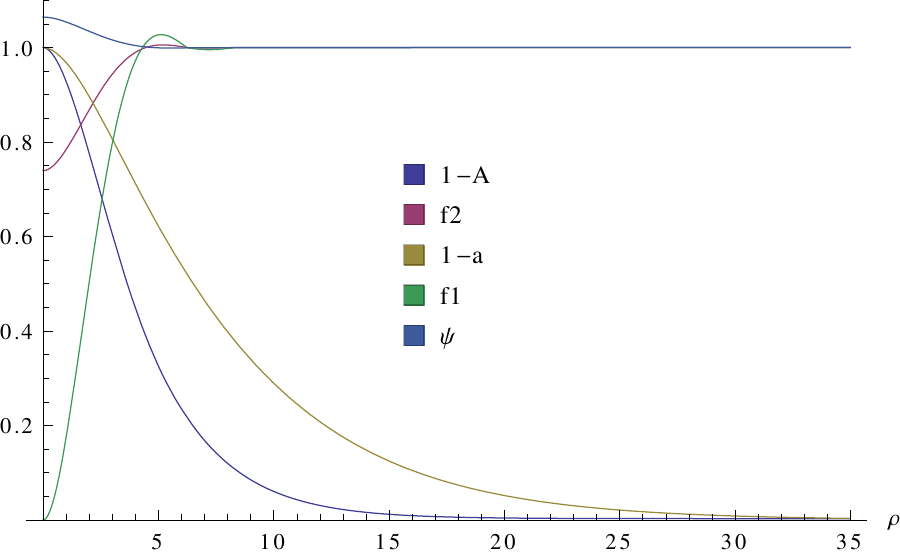}
\caption{
A numerrical solution for the profile functions of a global vortex in the second model.
The choice of parameters is:
$\lambda_g = 1, \lambda_1 = 0.01, \lambda_2 = 0.01, \lambda_e = 
1, \lambda_\psi = 5, m = 1, \xi = 2, e_{} = 0.05, g = 0.1$.}
\label{profile-2}
\end{figure}

Here we discuss the case where $\mu^{(2)} = 0$ and follow the same procedure we discussed before. 
Similarly to the analysis discussed in the previous section, 
there exists a global rotation of the doublet field to cancel the AB phase of the doublet. 
We add a single profile function for it as 
 \begin{eqnarray}
\label{phivortex3}
&&\Phi(r, \theta) = 
|\Phi|_v \left(
\begin{array}{ccc}
 0 & f_1(r) e^{i\theta}     \\
f_2(r)   & 0  \end{array}
\right),  \quad 
\Psi_v =\frac{ |\Psi|_v}{\sqrt 2}  \psi(r) \left(
\begin{array}{c}
1
 \\
1
\end{array}
\right)
\nonumber\\
&&
A_i(r, \varphi) = -\frac{\c}{4g} \frac{\epsilon_{ij}x_j }{r^2}\sigma^3A(r),\quad
a_i (r, \theta) = - \frac{\c}{2e_{}} \frac{\epsilon_{ij}x_j }{r^2}a(r).
\end{eqnarray}
The profile functions
$f_1(r), f_2(r), A(r), a(r)$ and $\psi(r)$ depend only on the radial coordinate $r$ 
with the boundary conditions
$f_1(0) = f_2'(0) = \psi(0) = 0,\quad f_1(\infty) = f_2(\infty) = \psi(\infty) = 1, A(0)= a(0) = 0, \quad A(\infty)= a(\infty) = 1$.
The constants $|\Phi|_v, |\Psi|_v$ and $\c$ have been defined before and they remain 
the same in this case too.  
 The equations of motion can be written as
\begin{eqnarray}
&&-  r\frac{\p}{\p r} \l(\frac{A'}{r}\r) + 4 g^2 |\Phi|_v^2  
             \l[\l( \frac{a + A}{2} - \frac{1}{\c}\r)f_1^2  + \frac{(A - a)}{2 }f_2^2\r] 
+  g^2|\Psi|_v^2  A \psi^2 = 0, \nonumber\\
&&-  r\frac{\p}{\p r} \l(\frac{a'}{r}\r) 
   + 4 e_{}^2|\Phi|_v^2  \l[\l( \frac{a + A}{2} - \frac{1}{\c}\r)f_1^2+ \frac{( a - A)}{2 }f_2^2\r]  = 0, \nonumber\\
&& - \frac{1}{r}\frac{\p}{\p r}(r f' ) +\l(1 - \c^2\frac{a + A}{2}\r)^2 \frac{f_1}{ r^2}  
 + |\Phi|_v^2\l[\lambda_g\l(f_1^2 - f_2^2\r)f_1
 + \frac{\lambda_e}2\l(f_1^2 + f_2^2 - 2\frac{\xi^2}{|\Phi|_v^2}\r)f_1 \r] \nonumber\\ 
&& \quad\quad 
   +  (\lambda_1 +  \lambda_2) |\Psi|_v^2 \psi^2 f_1 = 0, \nonumber\\ 
 &&   - \frac{1}{r}\frac{\p}{\p r}(r f_2' ) +  
\c^2\frac{(a - A)^2}{4 r^2}f_2 + |\Phi|_v^2\l[   \lambda_g\l(  f_2^2 - f_1^2\r)f_2
+  \frac{\lambda_e}2\l(f_1^2 + f_2^2 - 2\frac{\xi^2}{|\Phi|_v^2}\r)f_2 \r] 
\nonumber \\
&& \quad\quad
+ \lambda_1 |\Psi|_v^2 \psi^2  f_2 = 0, \nonumber\\ 
 && -  \frac{1}{r}\frac{\p}{\p r}(r \psi' ) 
 +  \frac{\c^2}{8 r^2}A^2\psi
 + \l[ m^2   + 2 \lambda_\psi |\Psi|_v^2 \psi^2 + \lambda_1 (f_1^2 + f_2^2) |\Phi|_v^2
 +  \lambda_2  f_1^2  |\Phi|_v^2 \r] \psi= 0. 
\end{eqnarray}
By solving these equations numerically, we obtain a solution in Fig.~\ref{profile-2}.
%%%%
%%%%%

%%%%%%%%%%%%
\subsection{The interactive case: an Alice string confined by a domain wall}
Since the hypercharge of the doublet is zero in this case, the AB phase of the doublet 
receives a contribution only from the non-Abelian gauge field 
around the Alice string  while encircling around the vortex, as in Eq.~(\ref{eq:AB2}).
The AB phase is a $\pi$ phase difference between $\psi_1$ and $\psi_2$, and  
so in this case too, the doublet looses its single-valuedness. 
Therefore, this system should create a domain wall around the vortex to recover single-valuedness of the doublet. To understand the existence of domain wall,  we add an another phase ($\phi$) to regain its 
single-valuedness similarly to the previous case. Here, we keep the Alice string away from the backreaction for simplicity.
We write the domain wall ansatz as
\begin{eqnarray}
\Psi_{\rm DW} \sim \frac{|\Psi|_v}{\sqrt 2}e^{i\frac{\theta + \phi(\theta)}{4}\sigma^3}
 \left(
\begin{array}{c}
1  \\
1
\end{array}
\right), \quad \phi(0) = 0, \quad \phi(2\pi) = -2\pi,  \label{eq:wall2}
\end{eqnarray}
with a decreasing function $\phi$.
Since the interaction term (the $\mu^{(2)}$ term) is quadratic in the triplet field, we should fix its magnitude so that it remains much lower 
than  the bare mass term of the triplet. We write the full potential (we set $\lambda_1 = \lambda_2 =0 $ for simplicity) and insert the doublet ansatz along with the Alice string configuration in Eq.~(\ref{eq:Alice-asymptotic})  
into the potential, to find 
\begin{eqnarray}
V(\Phi,\Psi) 
 &=&  2\lambda_e\left( |\Phi|_v^2 -  \xi^2\right)^2 + 
  \frac{\lambda_4}{4} [|\Psi|_v^2 - v_2^2]^2 + \frac{\mu^{(2)} }{2}|\Phi|_v^2|\Psi|_v^4\l(1 - \cos\phi\r)  .
\end{eqnarray}
From the above expression  we fix the coefficients with the relation $
 4\lambda_e \xi^2 > \mu |\Psi|_v^4$
to keep the quadratic term of the triplet always negative.
 Now we set $ |\Psi|_v = v_2,  |\Phi|_v = \xi$ and
the effective Hamiltonian is found to be 
\begin{eqnarray}
16\mathcal{ H}/v_2^2= \l(\p_i\phi\r)^2 +  2 m_0^2\l(1 - \cos\phi\r) , 
\quad m_0^2 \equiv 4 \mu^{(2)}\xi^2 v_2^2,
\end{eqnarray}
which is the sine-Gordon model, 
and a kink is inevitable from the boundary condition of $\phi$ in Eq.~(\ref{eq:wall2}). 
The solution is the well-known kink soliton
$\phi(x) = 4 \tan^{-1}e^{\pm  m_0 x}$
with the energy per area 
$T_{\rm DW} =  m_0 v_2^2 = 2 \sqrt{\mu^{(2)}} \xi v_2^3$. 
 In this case, 
the non-trivial element of $\mathbb{Z}_2$ appears as the AB phase and it is unavoidable 
since $\psi_1 = \pm \psi_2 \ne 0$ in the vacua.

%%%%%
\begin{figure}[htbp]
\centering
\begin{tabular}{ccc}
\includegraphics[totalheight=3.5cm]{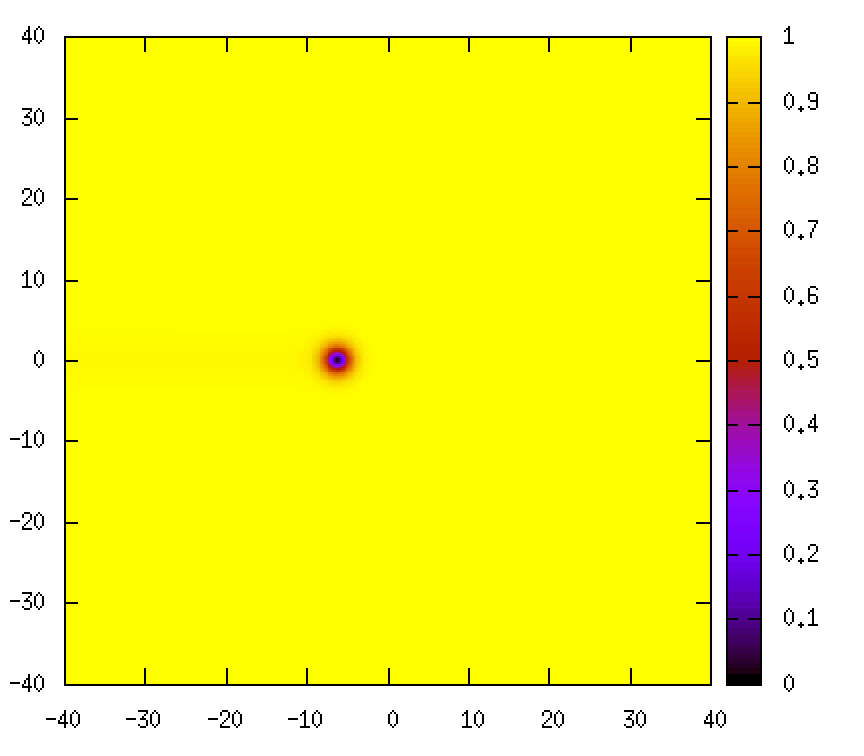} &
\includegraphics[totalheight=3.5cm]{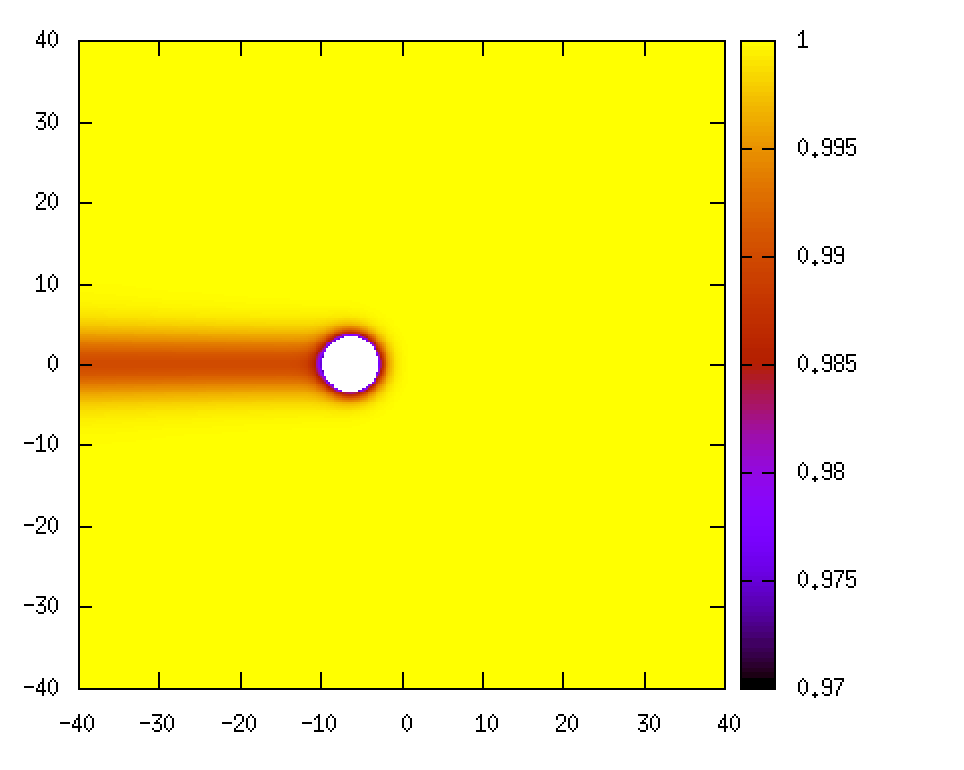} &
\includegraphics[totalheight=3.5cm]{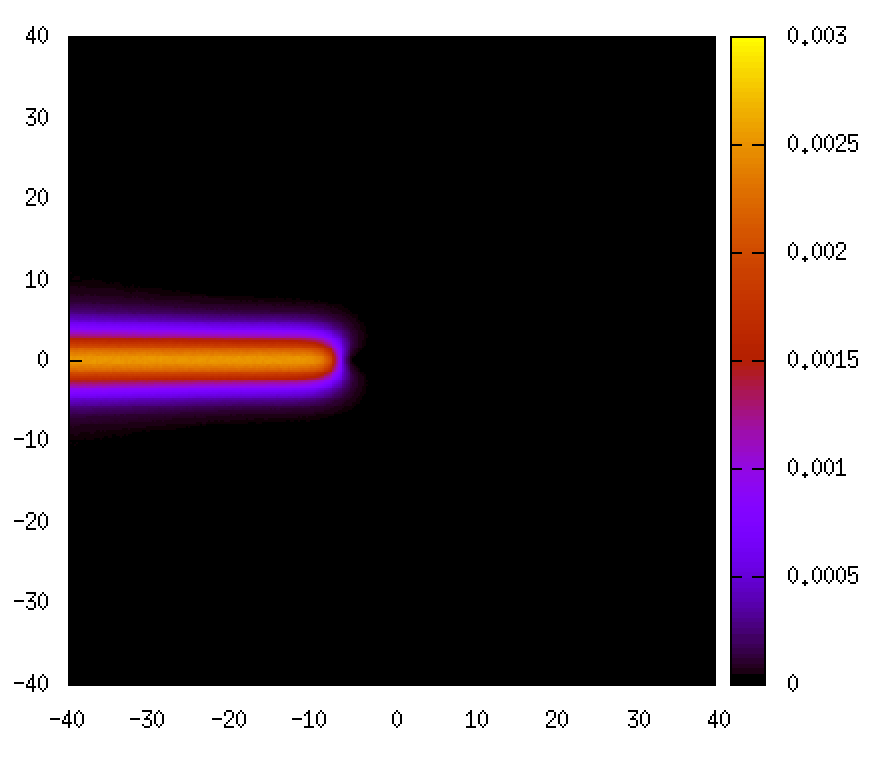}  \\
(a) & (b) & (c)\\
\includegraphics[totalheight=3.5cm]{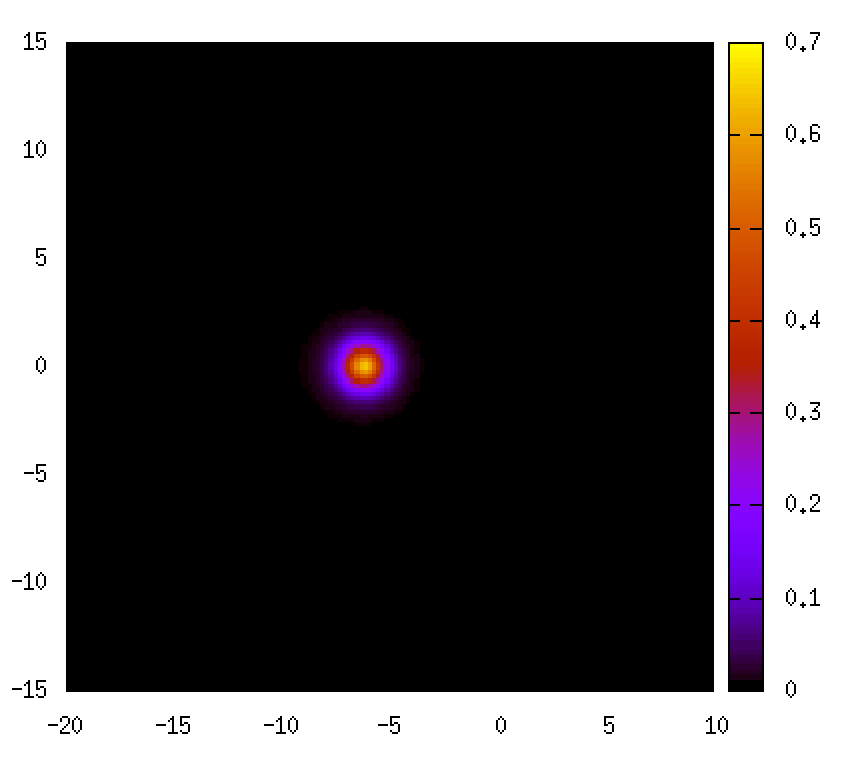} & 
\includegraphics[totalheight=3.5cm]{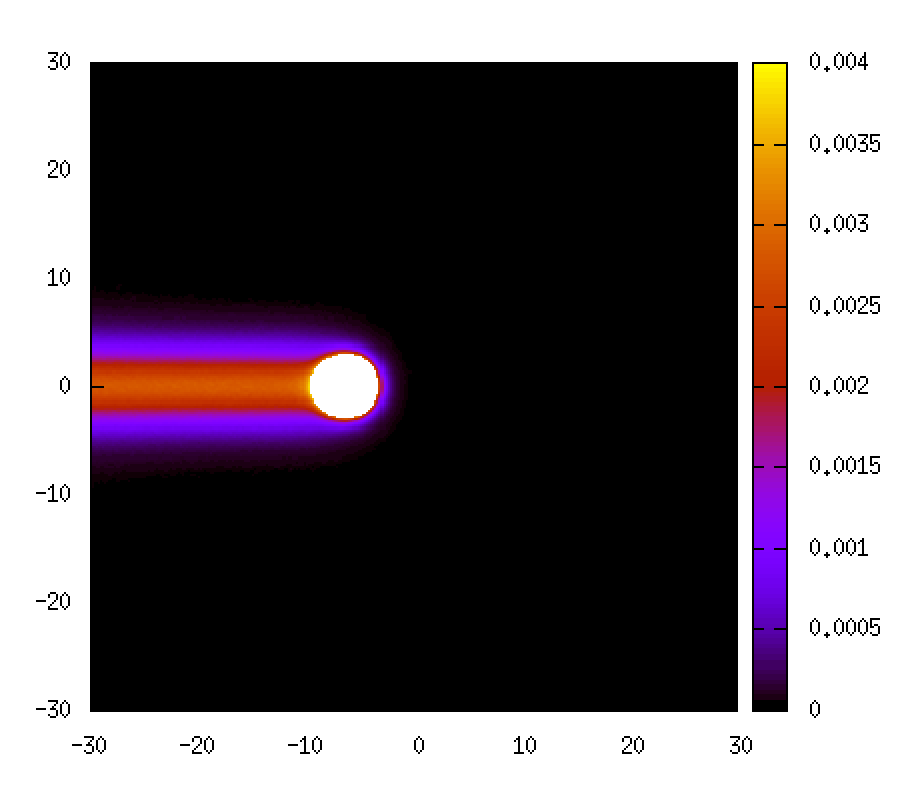} &
\includegraphics[totalheight=3.5cm]{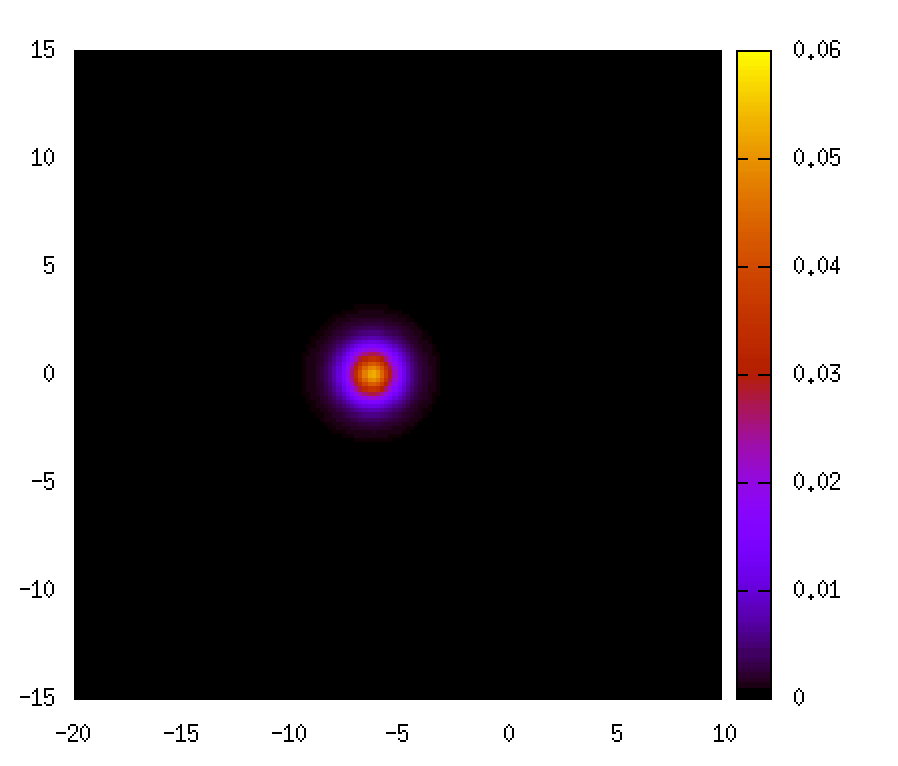}\\
(d) & (e) & (f)
\end{tabular}
\caption{An Alice string confined by a domain wall in the second model with a chargeless doublet. 
       The plots are the same with Fig.~\ref{DW_Alice1}.
       The parameter choices are  $\xi = 0.1,   e_{} = 0.5, g = 1,
      \mu^{(2)} = 100 , \lambda_\psi= 1, m =  0.1, \lambda_e= 2, \lambda_g=0.5 $. 
       (a)--(f) are the plots of the same quantities with Fig.~\ref{DW_Alice1}.
      }
\label{DW_Alice2}
\end{figure}

 To confirm our claim, we solve the full two-dimensional 
 equations of motion numerically by the relaxation method, 
 as shown in Fig.~\ref{DW_Alice2}. 
 The setting of numerics is the same with the first model.
 
 It is again interesting to note that the spatial rotation around the string is spontaneously broken due to 
 the interaction term proportional to $\mu^{(2)}$.

 %%%%%%%%%%%%%%%%
 \section{Summary and discussion}\label{sec:summary}
 In this paper, we have examined a question what happens when 
 fields receiving an AB phase develop a VEV. 
 To concretely study this problem,
we have considered 
an Alice string in the $SU(2) \times U(1)$ gauge theory with complex triplet scalar fields, 
and have introduced doublet scalar fields which are charged or chargeless for the $U(1)$ gauge group 
 in the first or second model, respectively. 
 The doublet scalar fields receive non-trivial AB phases when encircling 
 around the Alice string in the both models.
   We have found that, when the doublet develops a VEV, the Alice string turns to a global string with fractional flux for the both models 
   in the absence of the interaction depending on the relative phase between the doublet and triplet fields,  
   while the Alice string is attached by a soliton or a domain wall in the first or second model, respectively, 
   in the presence of such an interaction. 
   The interaction terms spontaneously break also the spatial rotation around the Alice string.
   For the both models, we have examined this first by showing that 
   the relative phase between the doublet and triplet fields is reduced to the sine-Gordon model
   at large distances from the Alice string, 
   and further have confirmed this by full 2D numerical simulations. 
  
Although a sine-Gordon kink is attached to the Alice string in the both models, 
the interpretation of the kink seems to be different in the both models.
It is a non-topological soliton in the first model 
while it is a topological domain wall connecting two disconnected vacua 
in the second model. 
This difference may be responsible for the stability of the soliton or domain wall.

The first model is related to the so-called triplet Higgs model proposed as one of the model beyond the SM, 
for which one adds a triplet Higgs scalar field in addition to the doublet Higgs scalar field of the SM. 
However, in the realistic case,  
 the VEV of the triplet Higgs scalar field should be much smaller than
the VEV of the doublet Higgs scalar field, to be consistent with the so-called $\rho$ parameter, 
while we have considered an inverse hierarchy in this paper.
In addition to this, the phases of the triplet fields are different for the two cases: 
a triplet VEV spontaneously brakes 
the $SU(2)_W \times U(1)_{Y}$ down to $U(1)_{W+Y}$ in the ferromagnetic phase, 
and to $O(2)$ for the polar phase. The former is relevant for the SM while 
the latter corresponds to our model admitting an Alice string. 
 The both cases fall into the same model but in different parameter regions. 
Although our parameter region is not realistic in the current Universe, 
it could be relevant in past in early Universe.

Although we have considered particular models admitting an Alice string, 
our conclusion seems to hold in more general cases. 
Namely, when fields receiving AB phases develop VEVs, 
a domain wall or soliton will be created in order for the fields to be single-valued,
or a string turns to a global string with a fractional flux. 
We call such a soliton or domain wall induced by AB phases as an ``AB defect.''
 
 One example is
the Georgi-Machacek model  \cite{Georgi:1985nv}, 
which was also proposed as a model beyond the SM, 
having three real triplet scalar fields in addition to the Higgs doublet.
In this model, the hierarchy of the VEVs of triplet and doublet Higgs fields 
can be interchanged consistent with the $\rho$ parameter.
If only the triplet fields develop a VEV at high energy, it admits a ${\mathbb Z}_2$ string, 
and if the doublet also develops a VEV at low energy, a domain wall is attached to it 
\cite{Chatterjee:2018znk}, similarly to our model.\footnote{
   As far as we understand, the two Higgs doublet model (2HDM), which is more popular extension of the SM, 
   is {\it not} an example of AB defects although it also admits a similar vortex-domain wall composite 
   \cite{Dvali:1993sg,Dvali:1994qf,Eto:2018hhg,Eto:2018tnk}. This is because one Higgs doublet allows no topological stable string, and instead the relative phase of the two Higgs doublets has a topological winding of vortex strings. In this sense, 2HDM is closer to axion models.
   } 
The difference with the current study is
that the ${\mathbb Z}_2$ string in that case is not an Alice string, 
and instead it is a non-Abelian string carrying ${\mathbb C}P^1$ moduli 
in the limit of the exact custodial symmetry.
The ${\mathbb C}P^1$ moduli of the domain wall and string match at the junction line.

Another example is given by two-gap (or two-component) superconductors,
which can be described by a $U(1)$ gauge theory with two charged complex scalars (gaps)
$\Phi_1$ and $\Phi_2$, coupled to each other by a Josephson term $\Phi_1^* \Phi_2 + {\rm c.c.}$
In fact, fractional fluxes were first found in this case \cite{Babaev:2001hv,Babaev:2004rm}.
Let us first assume the VEV $v_1 = \left<\Phi_1\right>$ only for the first component, 
and consider a vortex in the first component $\Phi_1$. Then, $\Phi_2$ receives the AB phase around the vortex. 
If the second component $\Phi_2$ develops a small VEV $v_2 = \left<\Phi_2\right>$ 
 at low energy with a hierarchy $v_1 \gg v_2$ between the VEVs, 
 there appears an AB defect attached to the vortex. 
A salient feature of this case is that the AB defect attached to the vortex in the first component 
$\Phi_1$
can end on a vortex in the second component $\Phi_2$, with the total configuration being a vortex molecule. 
This is because in this case the second component $\Phi_2$ also 
breaks the $U(1)$ gauge symmetry simultaneously, 
resulting in a nontrivial topology of $\pi_1$, in contrast to our Alice model and the GM model, 
in which the doublet does not allow a topologically stable vortex. 
In the case of two-gap superconductors, the hierarchy between the VEVs 
is not essential for the stability; 
either the case of $v_1 \gg v_2$ or $v_2 \gg v_1$ admits essentially the same vortex molecule.
In other words, 
the hierarchy of the VEVs is exchangeable.
On the other hand, in our Alice model and the GM model,
the hierarchy of the triplet VEV much larger than the doublet VEV allows a string attached by 
a soliton or domain wall as discussed in this paper. 
However, 
the inverse  hierarchy allows no topological string  
since only the doublet breaks the $SU(2) \times U(1)$ symmetry 
in the same way with the SM, 
allowing no topologically stable solitons; 
the hierarchy of VEVs are not exchangeable for the Alice model and GM model.
   
   Several more discussions are addressed here. 
   
In this paper, we have considered the case of only one Alice string. Our theory admits multiple BPS Alice strings at arbitrary positions, when we turn off the doublet field. Since a doublet encircling two strings receives no AB phase,  the two Alice strings are connected by one AB defect when the doublet develops a VEV. One natural question is how each Alice string find a partner, when there are many Alice strings.

  An Alice string carries a $U(1)$ modulus, corresponding to the internal direction of the flux 
  \cite{Chatterjee:2017hya}.
   When a $U(1)$ modulus is twisted along a closed Alice string (called a vorton), 
   it is nothing but a magnetic monopole 
  \cite{Shankar:1976un,Bais:2002ae,Striet:2003na}.
  See Ref.~\cite{Ruostekoski:2003qx} for a global analogues.
  It is a natural question what is the fatuous of the monopole if one introduces a doublet having a VEV. 
  In this case, a monopole as a twisted Alice ring may become a drum vorton \cite{Carter:2002te,Buckley:2002mx}, that is, a soliton or a wall is stretched inside the ring. A question may remain how magnetic fluxes from the monopole are confined.
   
   In Refs.~\cite{Sato:2018nqy,Chatterjee:2019rch}, Alice strings were applied to an axion model to solve the so-called domain wall problem.
$N$ axion domain walls are attached to one axion string in the model with a domain wall number $N$. In this case, domain walls cannot decay and dominate Universe, resulting in the domain wall problem. Although the domain wall number is two in the model in Refs.~\cite{Sato:2018nqy,Chatterjee:2019rch}, one axion string attached by two domain walls decays into two Alice strings each of which is attached by only one domain wall, thereby solving the domain wall problem. In Ref.~\cite{Sato:2018nqy}, the doublet scalar field was considered. This doublet field receives an AB phase as discussed in the present paper. Therefore, if the doublet develops a VEV, the two Alice strings as a result of the decay of one axion strings are connected by another domain wall (an AB defect) suppressing the decay. 
   
   It is also important to ask whether fermions receiving non-trivial AB phases can contribute to AB defects. 
   It depends on the form of fermion condensations;
   A fermion-anti-fermion condensation will have a trivial AB phase 
   while a fermion-fermion condensation will posses a non-trivial AB phase.
   Two-gap superconductors can be in fact regarded as such an example of 
   a gap composed of fermion-fermion condensation with a non-trivial AB phase.
   A non-Abelian vortex in dense QCD 
   \cite{Balachandran:2005ev,Nakano:2007dr,Eto:2009kg,Eto:2009bh,Eto:2009tr,Eto:2013hoa} 
   provides a non-trivial AB phase for charged particles
   \cite{Chatterjee:2015lbf} as well as a ${\mathbb Z}_3$ AB phase for quarks  
   \cite{Cherman:2018jir,Chatterjee:2018nxe,Chatterjee:2019tbz,Hirono:2018fjr},
   and so if (further) diquark condensation forms, it may give a fermion example of AB defects.
   
  It will be natural to ask whether the notion of AB defects can be extended to higher dimensional cases. 
  The example that we discussed in this paper is a string of codimension two attached by a wall of codimension one. If we consider one higher codimensions, a monopole can be attached by a string. In fact, many examples of such configurations are known 
  \cite{Tong:2003pz,Eto:2006pg,Auzzi:2003em,Shifman:2004dr,Hanany:2004ea,Eto:2004rz,
  Nitta:2010nd,Tong:2005un,Shifman:2007ce,Shifman:2009zz} 
  but it is unclear whether they can be understood in terms of generalized AB phases.
   To investigate this problem, the notion of topological obstructions of a monopole 
\cite{Nelson:1983bu,Abouelsaood:1982dz,Balachandran:1982gt} 
   may be useful. 
   
\section*{Acknowledgment}
This work is supported by the Ministry of Education, Culture, Sports, Science (MEXT)-Supported Program for the Strategic Research Foundation at Private Universities ``Topological Science" (Grant No. S1511006). 
This work is also supported in part by 
JSPS Grant-in-Aid for Scientific Research (KAKENHI Grant No.~19K14713 (C.~C.), No.~16H03984 (M.~N.), No.~18H01217 (M.~N.)), and also by MEXT KAKENHI Grant-in-Aid for Scientific Research on Innovative Areas ``Topological Materials Science'' No.~15H05855 (M.~N.).

 \appendix
\section{Numerical computation}\label{sec:appendix}
In this section we present full numerical results of the equations of motion. 
We use the relaxation method 
to find solutions. We use the static Hamiltonian with $\lambda_1 = \lambda_2 = 0$ 
for simplicity, given as
 \begin{eqnarray}
\label{h-num}
\mathcal{H} &=& \half\Tr F_{ij}^2 +  \frac{1}{4} f_{ij}^2 + |\D_i\Phi|^2 + |\D_i\Psi|^2 + V(\Phi, \Psi)\\
V(\Phi, \Psi) &=& V_\Phi (\Phi) + V_\Psi (\Psi) + V_{\rm int}(\Phi, \Psi) \\
 V_\Psi(\Phi) &=& \frac{\lambda_g}{4} \Tr[\Phi,\Phi^\dagger]^2 +  \frac{\lambda_e}{2}\left(\Tr \Phi\Phi^\dagger - 2 \xi^2\right)^2, \\
 V_\Phi(\Psi) &=&  M^2 \Psi^\dagger\Psi + \lambda_\psi \l(\Psi^\dagger\Psi\r)^2 .% 
\end{eqnarray}
The interaction $V_{\rm int}(\Phi, \Psi)$ is either
$V_{\rm int}^{(1)}$ and $V_{\rm int}^{(2)}$ for the first or second model:
\begin{eqnarray}
 V_{\rm int}^{(1)}(\Phi, \Psi) &=& \mu^{(1)}  \l( \Psi_c^\dagger \Phi^* \Psi +   \Psi^\dagger \Phi \Psi_c\r),
 \quad
 V_{\rm int}^{(2)}(\Phi, \Psi) = \mu^{(2)} \l| \Psi_c^\dagger \Phi^* \Psi\r|^2  . 
\end{eqnarray}

For the relaxation method, 
we use the equations of motion of the first order in imaginary time as
\begin{eqnarray}
\frac{\p \Phi^\a}{\p \tau} &=& - \frac{\delta \mathcal{H}}{\delta {\Phi^*}^\a},\\
\frac{\p \Psi^\a}{\p \tau} &=& - \frac{\delta \mathcal{H}}{\delta {\Psi^*}^\a}\\
\frac{\p A_i^\a}{\p \tau} &=& - \frac{\delta \mathcal{H}}{\delta A_i^\a}\\
\frac{\p a_i}{\p \tau} &=& - \frac{\delta \mathcal{H}}{\delta a_i}
\end{eqnarray}

We use a $500\times500$ square lattice with a lattice spacing $0.2$, 
subjected by the Neumann boundary conditions.

The initial configurations used in these computations are
\begin{eqnarray}
  \Phi^1(x,y) &=& 2 \xi \tanh(\sqrt{(x^2 + y^2)}) \exp( i 0.5 n_v \theta)\cos(0.5 n_v \theta)),\\
   \Phi^2(x,y) &=& - 2 \xi \tanh(\sqrt{(x^2 + y^2)}) \exp( i 0.5 n_v \theta)\sin(0.5 n_v \theta)),\\
    \Phi^3(x,y) &=& 0,\\
    a_1(x, y)&=& - (0.5/e_{})n_v \frac{y}{L^2}  \tanh(\sqrt{(x^2 + y^2)}),\\
       a_2(x,y)&=& (0.5/e_{}) n_v \frac{x}{L^2} \tanh(\sqrt{(x^2 + y^2)}),\\
       A^1_1(x,y)&=& A^1_2(x,y) =0 ,\\
       A^2_1(x,y) &=& A^2_2(x,y) = 0 ,\\
        A^3_1(x,y) &=& - (0.5/g)n_v \frac{y}{L^2}  \tanh(\sqrt{(x^2 + y^2)}),\\
       A^3_2(x,y)) &=&  (0.5/g) n_v \frac{x}{L^2} \tanh(\sqrt{(x^2 + y^2)}),\\
       \psi_1(x, y) &=& \sqrt{\frac{-0.5 M^2}{\lambda_\psi}},\\
        \psi_2(x, y) &=& 0.
\end{eqnarray}
Here $n_v(=1)$ is a winding number and $L$ is the system size.

%%%%%%%%%%%%%%%%%%%%%%%%%%%%%%%%%%%%%%%

\end{document}